\documentclass[reprint,prl,amssymb,twocolumn,supersriptaddress,groupeaddress]{revtex4-1}
\usepackage{amssymb}
\usepackage{amsmath}
\usepackage{graphicx}
\usepackage{dcolumn}
\usepackage{bm}
\usepackage{bm}
\usepackage{natbib,hyperref}
\usepackage{hyperref}
\usepackage{color}
\hypersetup{colorlinks=true, 
    linkcolor=red,          
    citecolor=magenta,        
    filecolor=gree,      
    urlcolor=blue           
}

\begin{document}


\title{Amplitude tuning of steady state entanglement in strongly driven coupled qubits}

\author{Ana Laura Gramajo, Daniel Dom\'{\i}nguez and Mar\'{\i}a Jos\'{e} S\'{a}nchez}

\affiliation{Centro At{\'{o}}mico Bariloche and Instituto Balseiro, 8400 San Carlos de Bariloche, Argentina.\\
}




\date{\today}%
\begin{abstract}
In this work we report on a new mechanism to generate dissipative steady state entanglement in two coupled qubits driven by strong periodic ac fields.
We show that steady entanglement can be generated at one side of a multiphoton resonance between a non-entangled ground state and an entangled excited state. The degree of entanglement can be tuned as a function of the amplitude of the periodic drive. 
A rich dynamic behavior with {\it creation, death and revival} of entanglement can be observed for certain parameter regimes, accessible in current experimental devices.

\end{abstract}
\maketitle

The generation and stabilization of entanglement is  one of the main challenges  in quantum information applications. 
In recent years strategies  based on the creation of steady state entanglement through engineered dissipation have been discussed theoretically \cite{kraus_2008,verstraete_2009,reiter_2013} and demonstrated in experiments \cite{barreiro_2011,lin_2013,kienzler_2015,krauter_2011,shankar_2013,leghtas_2013,kimchi_2016}.
In this scheme, the  system  of interest is driven by external fields
and   coupled to a reservoir, developing    a  nontrivial  non-equilibrium dynamics  that leads to a highly entangled steady state. The effective relaxation rates can be  tuned by adequately
designing  the quantum reservoir,  the system-reservoir couplings or the driving protocols.
Experimental demonstrations include realizations with trapped ions \cite{barreiro_2011,lin_2013,kienzler_2015}, atomic
ensembles \cite{krauter_2011}, and superconducting qubits \cite{shankar_2013,leghtas_2013,kimchi_2016}.
Another strategy for entanglement stabilization are measurement based protocols, which have been implemented, for example, in coupled superconducting qubits
\cite{dicarlo_2009,riste_2013,roch_2014,chantasri_2016,liu_2016}.

The different  proposed mechanisms for driven dissipative entanglement generation
utilize  {\it weak} resonant drivings to tailor the  relaxation processes \cite{kraus_2008,verstraete_2009,reiter_2013,barreiro_2011,lin_2013,kienzler_2015,krauter_2011,shankar_2013,leghtas_2013,kimchi_2016}.  However, for {\it large} amplitude  periodic drivings, interesting non perturbative effects are known to exist. Among these, 
 coherent destruction of tunneling \cite{grossmann_1991,bloch_2012,gagnon_2017},  Landau-Zener-St\"uckelberg (LZS) interferometry \cite{oliver_2005,sillanpaa_2006,berns_2006,rudner_2008,izmalkov_2008,shevchenko_2010,wilson_2010,dupont_2013,forster_2014,neilinger_2016} and bath-mediated population inversion \cite{stace_2005,stace_2013,ferron_2012,ferron_2016}  have been studied in two-level systems.

Relying on these later effects,  we  present a new mechanism to induce steady state entanglement. 
Using as a test system two coupled qubits, we will demonstrate that the entanglement in the steady state 
can be induced and tuned by changing the amplitude of a driving periodic field.  One of our main results is advanced in Fig.1(a) where we show how the concurrence (a measure of entanglement) can be increased or decreased as a function of the amplitude of the periodic driving.

\begin{figure}[!htb]
	\includegraphics[scale = 0.38]{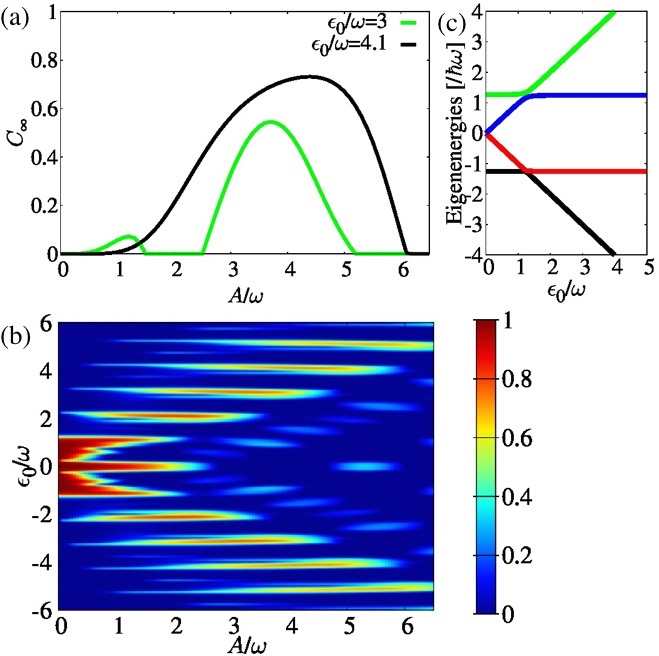}
	\caption{(a) Plots of the steady state concurrence, $C_{\infty}$, as function of  the driving amplitude $A/\omega$  for $\epsilon_{0}/\omega = 3$ (green line) and $\epsilon_{0}/\omega = 4.1$ (black line). (b) Colour map of  $C_{\infty}$ versus $A/\omega$ and $\epsilon_{0}/\omega$. (c) Eigenenergies $E_{i}$ of the system Hamiltonian $H_0$  as a function of $\epsilon_{0}/\omega$. Along this work, we choose  $\Delta_{2}/\Delta_1=1.5$ , $J/\Delta_1=-25$ and $\omega/\Delta_1=10$. The bath temperature is taken as $T_{b}/\Delta_1=0.0467$ and for the bath spectral density we consider $\gamma=0.001$ with a cutoff frequency $\omega_c/\Delta_1=333$. See text for details. }
\label{fig:1}
\end{figure}

 In this work  we consider two coupled qubits   with Hamiltonian  $\hat{H}_s(t)=\hat{H}_{0} + \hat{V}(t)$, where
 \begin{equation}
 \hat{H}_{0} = \sum^{2}_{i=1}\left(-\frac{\epsilon_{0}}{2}\sigma_{z}^{(i)} - \frac{\Delta_{i}}{2}\sigma_{x}^{(i)}\right) -\frac{J}{2}\left(\sigma^{(1)}_{+}\sigma^{(2)}_{-} + \sigma^{(1)}_{-}\sigma^{(2)}_{+}\right) 
 \nonumber
 \end{equation}
  with  $\sigma^{(i)}_{z,x,+,-}$ the Pauli matrices in the Hilbert space of qubit $i$ \cite{nota}.
This type of Hamiltonian can be realized, for instance, in superconducting qubits \cite{berkley_2003,izmalkov_2004,majer_2005,liu_2006,zhang_2009,weber_2017},  where $\Delta_1,\Delta_2,J$ are fixed device parameters and $\epsilon_{0}$ can be controlled experimentally.
 The  external ac driving  field is $\hat{V}(t)=-A\cos(\omega t)\left(\sigma^{(1)}_{z} + \sigma^{(2)}_{z} \right)/2$, of amplitude $A$  and  frequency $\omega$ \cite{shevchenko_2008,ilichev_2010,satanin_2012,temchenko_2011,sauer_2012, gramajo_2017}. 
 
 Dissipation and decoherence are taken into account by considering the open system dynamics, with global Hamiltonian $\hat{\cal H}(t)=\hat{H}_{s}(t) + \hat{H}_{b} + \hat{H}_{sb}$.  The Hamiltonian of the bath is $\hat{H}_{b}$ corresponding to a system of independent oscillators. We consider a linear and weak system-bath coupling represented by the Hamiltonian $\hat{H}_{sb}=g\left(\sigma^{(1)}_{z} + \sigma^{(2)}_{z} \right)\otimes\mathcal{B}$, with  $\mathcal{B}$ an observable of the bath.
 We assume an Ohmic spectral density  $J(\Omega)=\gamma\Omega e^{-|\Omega|/\omega_{c}}$,  where $\gamma\propto g^2$, 
 and the bath is at equilibrium at temperature $T_b$.
We focus on the dynamics of the reduced density matrix $\rho(t)=\textit{Tr}_{b}\left(\rho_{tot}\right)$, obtained by tracing out   from the global density matrix $\rho_{tot}$,   the degrees of freedom of the thermal bath. We  solve the corresponding Quantum Master Equation under the Floquet-Born-Markov approach \cite{kohler_1997}, which allows for the treatment of driving forces of arbitrary strength and frequency in open systems. We calculate numerically the steady state $\rho(t\rightarrow\infty)$  and
the time dependent $\rho(t)$ taking as  initial condition the ground state of $\hat{H}_{0}$  (see Supplementary Information).

We choose as an entanglement measure the concurrence, which can be calculated for mixed states as $C=\text{max}\{ 0, \lambda_{4}-\lambda_{3}-\lambda_{2}-\lambda_{1} \}$,  where $\lambda_{i}$'s are real numbers in decreasing order and correspond to the eingenvalues of the matrix $R=\sqrt{\sqrt{\rho}\tilde{\rho}\sqrt{\rho}}$, with $\tilde{\rho}=\sigma^{(1)}_{y}\otimes\sigma^{(2)}_{y} \rho^{*}\sigma^{(1)}_{y}\otimes\sigma^{(2)}_{y}$ \cite{wootters_1998}. 

The manipulation  of entanglement  by an ac drive has been already studied in closed systems, neglecting  the effect of the thermal bath. For two isolated coupled qubits, the generation of entanglement  can occur at and near $n$-photon resonances  \cite{sauer_2012,gramajo_2017}, when the resonance is among two separable, {\it i.e.} disentangled, eigenstates of $H_0$ (SS resonance)  or when the resonance is among a separable eigenstate (taken as the initial condition) and an entangled eigenstate (SE resonance).
The system Hamiltonian $H_0$, for $\Delta_1,\Delta_2\ll\epsilon_0$, has two entangled eigenstates $|e_{\pm}\rangle\approx \frac{1}{\sqrt{2}}(|01\rangle \pm|10\rangle)$ (in the basis spanned by the eigenstates of  $\sigma^{(1)}_{z}\otimes\sigma^{(2)}_{z}$) with eigenenergies $E_{e\pm}\approx\mp J/2$, and two separable (disentangled) eigenstates $|s_0\rangle\approx|00\rangle$ and  $|s_1\rangle\approx|11\rangle$, with eigenenergies $E_{s0}\approx-\epsilon_0$ and $E_{s1}\approx\epsilon_0$, respectively.  The ground state is entangled ($|E_0\rangle\approx|e_-\rangle$) with concurrence $C\approx1$ for $|\epsilon_0|<|J|/2$ and separable ($|E_0\rangle\approx|s_0\rangle$)  for $|\epsilon_0|>|J|/2$,  with $C\approx0$. In Fig.1(c) we  plot the eigenenergies $E_i$ as a function of $\epsilon_0$ for $J=-25\Delta_1$.  Considering the condition $E_i-E_j\approx n\omega$, the SS resonances are for $2\epsilon_0\approx n\omega$ and the SE resonances for $\epsilon_0\pm J/2\approx n\omega$. (See the Supplementary Information.)

Let us analyze what happens for the open system situation considered in the present work.  From now on we will focus 
on the possibility of entanglement generation when the ground state is separable, $|\epsilon_0|>|J|/2$. 
Fig.\ref{fig:1}(b) shows the concurrence $C_{\infty}$ in the stationary regime, as a function of $A/\omega$ and $\epsilon_{0}/\omega$, for $J/\omega=-2.5$.
When the driving is on, $A\not=0$, we find that entanglement  generation takes place
 for certain values of $\epsilon_0$, which are close to the SE resonances at $\epsilon_0\approx|J|/2+n\omega$. As it is shown in detail in Fig.1(a), 
in these cases the concurrence is modulated by the driving amplitude and, by adequately tuning $A$,  $C_{\infty}$ can reach values close to 1, corresponding to a maximally entangled Bell's state, even when the ground state is separable.
\begin{figure}[t]
\includegraphics[scale = 0.38]{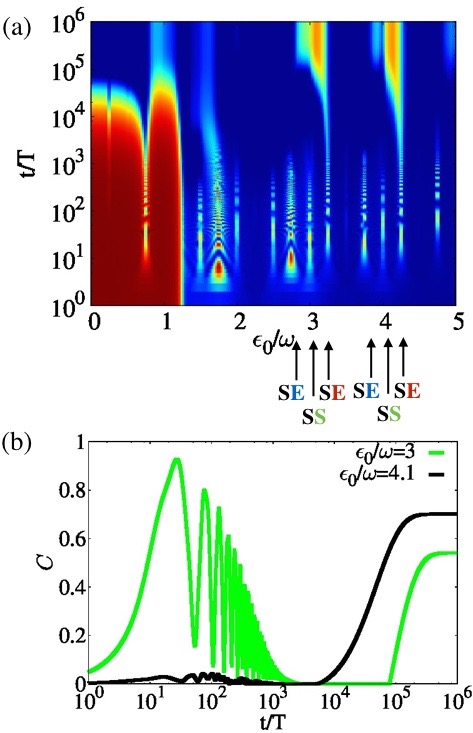}
\caption{(a) Intensity plot of  $C$ versus $\epsilon_{0}/\omega$ and $t/T$. The  location  of some of the SE(SS)-resonances are indicated. (b) Plot of $C$ as a function of the time $t/T$ for $\epsilon_{0}/\omega = 3$ (green line) and $\epsilon_{0}/\omega = 4.1$ (black line).  The initial condition corresponds to the ground state  for the correspondent $\epsilon_{0}/\omega$. The results correspond to $A/\omega=3.8$. Other parameters are the same as in Fig.\ref{fig:1}.} 
\label{fig:2}
\end{figure}

To understand why there is generation of steady entanglement near  SE resonances, it is necessary   to analyze in detail  the time-evolution of the system. 
 Fig.\ref{fig:2}(a) shows  $C(t)$ as a function of $\epsilon_{0}/\omega$ and the normalized time $t/T$ for a fixed value of amplitude $A/\omega=3.8$, with $T=2\pi/\omega$. It is straightforward to observe that the concurrence displays a rich dynamics near the multiphoton resonances. At short times there is a driving induced generation of entanglement at and near the values of $\epsilon_0$  corresponding to SS or to SE resonances. 
 This { \it short time dynamic entanglement creation}  is carried out by the coherent superposition of states induced by the driving, and  corresponds to the usual Rabi-like oscillations 
 at multiphoton resonances \cite{nakamura_2001,shevchenko_2010}.
 Similar results have been obtained for the isolated system, as we already mentioned \cite{gramajo_2017}.  In that case, the concurrence in the parameter space $[\epsilon_{0},A]$  presents a  pattern 
 that can be understood in terms of Landau-Zener-St\"uckelberg (LZS) interference, extensively studied and observed in single superconducting qubits \cite{shevchenko_2010}.
However, for times above the decoherence time, $t>t_c$ [with $t_{c}\sim 10^{3} T$ in Fig.2] , we find that the driven induced entanglement fades away in the case of the SS resonances.  In this situation, the entanglement  is fragile against the noise of the external environment and it is easily destroyed beyond the decoherence time. 

A strikingly different behavior takes place  in the case of SE resonances. At large time scales, above the relaxation time 
$t_{r} > t_c$, we find  the generation of steady entanglement at one  side of the SE resonances. As an example, we show in Fig.\ref{fig:2}(b) the time evolution of the concurrence for two off-resonant cases that are close to an SE resonance. For $\epsilon_0/\omega=4.1$ (shown in black line), that is below the SE resonance at $\epsilon_0/\omega=4.25$, we see that at initial times the entanglement is negligible (the concurrence is very small) and only after driving the system for large times, above $t_r\sim 10^4T$, steady entanglement is created. The entanglement induced in this later case is robust and stable at long times, opposite  to  the SS resonance situation previously described.
Another interesting and non trivial behaviour  takes place for $\epsilon_0/\omega=3$, which corresponds to an SS resonance that is very close to the SE resonance at $\epsilon_0/\omega=3.25$ (this $C(t)$ is plotted with a green line in Fig.\ref{fig:2}(b)). 
At $t=0$ the ground state is disentangled and $C\approx0$. After the driving is turned on, there is a dynamic generation of entanglement due to a  Rabi-like resonance among two separable states, giving place to  an oscillating $C(t)$ that can reach values close to $1$. At the decoherence time, this entanglement dies off and the concurrence drops to zero for $t>t_c\sim 10^3T$, and stays at this value for times up to $10^5 T$. Above this later time, steady state entanglement sets in, which is induced due to the nearness to the SE resonance at  $\epsilon_0/\omega=3.25$. Thus, those cases where SS and 
SE resonances are close, exhibit a rich  behavior as a function of time with {\it creation, death and revival} of entanglement.

\begin{figure}[!htb]
\includegraphics[scale = 0.32]{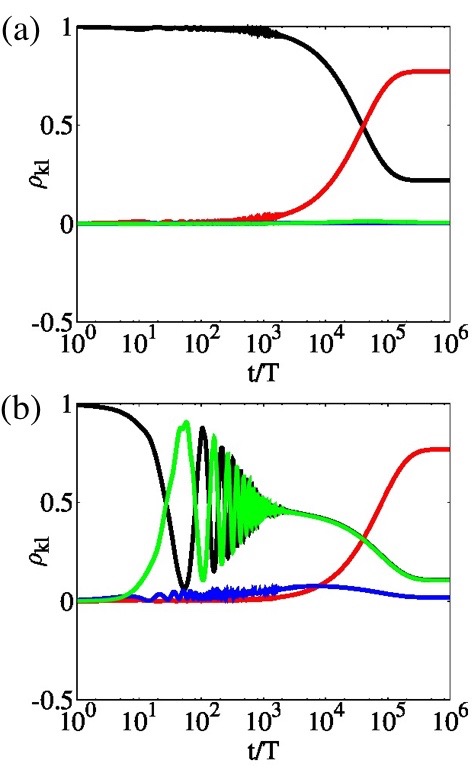}
\caption{Plots corresponding to the quantum tomography of two-qubits density matrix $\rho$ as function of normalized time $t/T$ for  $A/\omega=3.8$, and for $\epsilon_{0}/\omega=4.1$(a) and $\epsilon_{0}/\omega=3$(b). Matrix elements  $\rho_{kl}$ are in the eigenstates basis. The  colors of the lines correspond to the colors representing eigenstates in Fig.1(c) .}
\label{fig:3}
\end{figure}

The dynamics of the  two paradigmatic  examples discussed above can be better described in terms of  quantum tomography, by evaluating the time evolution of individual components of the density matrix. In Fig.\ref{fig:3} we present the plot of the density matrix elements $\rho_{kl}$ as function of $t/T$  using the eigenstates basis $\{|E_{k}\rangle\}$, with $k,l=0,...,3$. [See Fig.\ref{fig:1}(c), where the eigenstates are plotted versus $\epsilon_{0}/\omega$]. As the off-diagonal $\rho_{kl}$  (not shown in the plot)  become negligibly small above the decoherence time, the  interesting behavior is obtained for the populations, given by the diagonal terms $\rho_{ii}$. 
 Fig.\ref{fig:3}(a) shows the case  for $\epsilon_0/\omega=4.1$, where the ground state (black line) is separable, and is  close to a resonance with the first excited state (red line), which is entangled. The population of the two other eigenstates is negligible along all the time evolution, and the dynamics can be reduced to the subspace of the two states that are near resonance. Since the system is off-resonance, the population remains mostly in the ground state, which corresponds to the initial condition. At large times, above $t_r$, the population of the first excited state rapidly increases and the ground state is depopulated. This explains the sudden creation of entanglement shown in Fig.\ref{fig:2}(b) for this case, since the first excited state is entangled.
The case of creation, death and revival of entanglement is plot in Fig.\ref{fig:3}(b),  for  $\epsilon_0/\omega=3$.
Here, the ground state (black line) is at resonance with the third excited state (green line), and both are separable states. For times $t<t_{c}$, their populations display  Rabi-like oscillations while the populations of the other two states are negligible. Close to the decoherence time the oscillations are damped, and both populations tend to be equal to $1/2$. Above $t_{c}$ the coherence between these two states is lost, and  the concurrence vanishes. At larger time scales, above $t_r$,  a rapid transfer of population to the first excited state (red line) sets in, with almost all of the population being transferred to this entangled state. 


The behavior seen in  Fig.\ref{fig:3}(a), is reminiscent of the dynamic transition 
found in driven dissipative two-level systems near a multiphoton resonance \cite{ferron_2012,ferron_2016}, where population 
inversion can be  induced in the steady state.
The relaxation rate is strongly dependent on the amplitude $A$ and, in the case of two-level systems, it has been written as a sum of terms
$\Gamma_r(A)=\sum_n \Gamma_r^{(n)}(A)$ \cite{wilson_2010,grifoni_2010,ferron_2016}.
The $n$-th term can be interpreted as the contribution of $n$ virtual $\hbar\omega$ photons.
The $\Gamma_r^{(0)}$ corresponds to the direct relaxation from the excited state to the ground state, and it  is the dominant relaxation mechanism for low $A$.
 For large amplitudes, $A\gtrsim \omega$, the rates $\Gamma_r^{(n)}(A)$  oscillate as a function of $A/\omega$ with a Bessel-like dependence. 
 The dynamic transition leading to population inversion at the side of a $m$-photon resonance happens for the ranges of amplitude satisfying $\Gamma_r^{(-m)}(A)\gg\Gamma_r^{(0)}(A)$ \cite{ferron_2016}. In this case, the $\Gamma_r^{(-m)}$ term corresponding to  the absorption of $m$ photons from the ground state followed by a relaxation to the  excited state, prevails instead of the standard relaxation from the excited state to the ground state.

Near a $m$-photon resonance we can effectively concentrate on the subspace spanned by  the two  states intervening in the resonance. Within this approximation, we have calculated the rates   $\Gamma_r^{(n)}(A)$ (see the Supplementary Information). In Fig.\ref{fig:4}(a) we plot the time dependence of the concurrence as a function of $A/\omega$ for $\epsilon_0=4.1$ below a $m=3$-photon SE resonance. Here we see that for certain values of $A$ there is generation of entanglement in the steady state. Fig.\ref{fig:4}(b) shows the calculated $\Gamma_r^{(0)}(A)$ and $\Gamma_r^{(-3)}(A)$ rates as a function of $A/\omega$ while Fig.\ref{fig:4}(c) shows schematically the relaxation processes that correspond to each case. 
Comparing Figs.\ref{fig:4}(a) and (b) we see that steady state entanglement corresponds to the range of $A$ where $\Gamma_r^{(-3)}(A)\gg\Gamma_r^{(0)}(A)$. This shows that by tuning the value of $A$ near an SE resonance one can attain the conditions for populating the first excited state at long times, leading to the generation of stable steady state entanglement.

To summarize we have found  three different dynamical regimes for entanglement evolution in driven  coupled qubits: (i) Below the decoherence time, $t<t_c$, there is a dynamic generation of entanglement at multiphoton resonances, as described in \cite{sauer_2012,gramajo_2017}. (ii) For times $t_c<t<t_r$, there is a long time interval of {\it entanglement blackout}, where entanglement is destroyed due to decoherence with the environment. (iii)  Above the relaxation time, $t>t_r$, entanglement is created and preserved for long times  near the SE resonances. This later effect enables the generation of steady state  entanglement, which can be tuned as a function of the driving amplitude $A$. Quantum state tomography measurements \cite{roch_2014} in solid state devices where Landau-Zener-St\"{u}ckelberg interferometry has been studied in single qubits \cite{oliver_2005,sillanpaa_2006,berns_2006,rudner_2008,izmalkov_2008,shevchenko_2010,wilson_2010,dupont_2013,forster_2014,neilinger_2016} are good candidates to test this new mechanism for entanglement generation.

\begin{figure}[!htb]
\centering
\includegraphics[scale = 0.29]{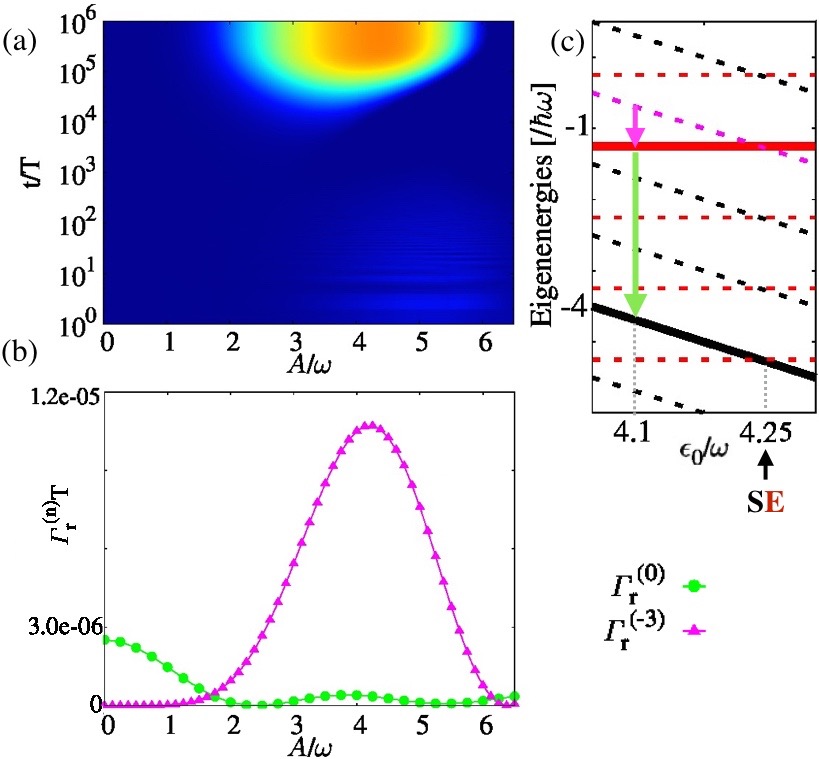}
\caption{(a) Intensity plot of  $C(t)$ as a function of $A/\omega$ and $t/T$. (b) Terms $\Gamma_{r}^{(n)}$ that contribute to the relaxation rate are plotted as a function of the driving amplitude $A/\omega$. Both cases correspond to  $\epsilon_{0}/\omega=4.1$. (c) Schematic representation of the involved transition processes between the states $|E_{0}\rangle$ (black line) and $|E_{1}\rangle$ (red line). The dashed lines represent the eigenenergies plus the addition of integer multiples of $\hbar\omega$.}
\label{fig:4}
\end{figure}


We acknowledge support from CNEA, CONICET (PIP11220150100756), UNCuyo (P 06/C455) and ANPCyT (PICT2014-1382, PICT2016-0791).


\begin{thebibliography}{49}%
	\makeatletter
	\providecommand \@ifxundefined [1]{%
		\@ifx{#1\undefined}
	}%
	\providecommand \@ifnum [1]{%
		\ifnum #1\expandafter \@firstoftwo
		\else \expandafter \@secondoftwo
		\fi
	}%
	\providecommand \@ifx [1]{%
		\ifx #1\expandafter \@firstoftwo
		\else \expandafter \@secondoftwo
		\fi
	}%
	\providecommand \natexlab [1]{#1}%
	\providecommand \enquote  [1]{``#1''}%
	\providecommand \bibnamefont  [1]{#1}%
	\providecommand \bibfnamefont [1]{#1}%
	\providecommand \citenamefont [1]{#1}%
	\providecommand \href@noop [0]{\@secondoftwo}%
	\providecommand \href [0]{\begingroup \@sanitize@url \@href}%
	\providecommand \@href[1]{\@@startlink{#1}\@@href}%
	\providecommand \@@href[1]{\endgroup#1\@@endlink}%
	\providecommand \@sanitize@url [0]{\catcode `\\12\catcode `\$12\catcode
		`\&12\catcode `\#12\catcode `\^12\catcode `\_12\catcode `\%12\relax}%
	\providecommand \@@startlink[1]{}%
	\providecommand \@@endlink[0]{}%
	\providecommand \url  [0]{\begingroup\@sanitize@url \@url }%
	\providecommand \@url [1]{\endgroup\@href {#1}{\urlprefix }}%
	\providecommand \urlprefix  [0]{URL }%
	\providecommand \Eprint [0]{\href }%
	\providecommand \doibase [0]{http://dx.doi.org/}%
	\providecommand \selectlanguage [0]{\@gobble}%
	\providecommand \bibinfo  [0]{\@secondoftwo}%
	\providecommand \bibfield  [0]{\@secondoftwo}%
	\providecommand \translation [1]{[#1]}%
	\providecommand \BibitemOpen [0]{}%
	\providecommand \bibitemStop [0]{}%
	\providecommand \bibitemNoStop [0]{.\EOS\space}%
	\providecommand \EOS [0]{\spacefactor3000\relax}%
	\providecommand \BibitemShut  [1]{\csname bibitem#1\endcsname}%
	\let\auto@bib@innerbib\@empty
	\bibitem [{\citenamefont {Kraus}\ \emph {et~al.}(2008)\citenamefont {Kraus},
		\citenamefont {B\"uchler}, \citenamefont {Diehl}, \citenamefont {Kantian},
		\citenamefont {Micheli},\ and\ \citenamefont {Zoller}}]{kraus_2008}%
	\BibitemOpen
	\bibfield  {author} {\bibinfo {author} {\bibfnamefont {B.}~\bibnamefont
			{Kraus}}, \bibinfo {author} {\bibfnamefont {H.~P.}\ \bibnamefont
			{B\"uchler}}, \bibinfo {author} {\bibfnamefont {S.}~\bibnamefont {Diehl}},
		\bibinfo {author} {\bibfnamefont {A.}~\bibnamefont {Kantian}}, \bibinfo
		{author} {\bibfnamefont {A.}~\bibnamefont {Micheli}}, \ and\ \bibinfo
		{author} {\bibfnamefont {P.}~\bibnamefont {Zoller}},\ }\href {\doibase
		10.1103/PhysRevA.78.042307} {\bibfield  {journal} {\bibinfo  {journal} {Phys.
				Rev. A}\ }\textbf {\bibinfo {volume} {78}},\ \bibinfo {pages} {042307}
		(\bibinfo {year} {2008})}\BibitemShut {NoStop}%
	\bibitem [{\citenamefont {Verstraete}\ \emph {et~al.}(2009)\citenamefont
		{Verstraete}, \citenamefont {Wolf},\ and\ \citenamefont
		{Ignacio~Cirac}}]{verstraete_2009}%
	\BibitemOpen
	\bibfield  {author} {\bibinfo {author} {\bibfnamefont {F.}~\bibnamefont
			{Verstraete}}, \bibinfo {author} {\bibfnamefont {M.~M.}\ \bibnamefont
			{Wolf}}, \ and\ \bibinfo {author} {\bibfnamefont {J.}~\bibnamefont
			{Ignacio~Cirac}},\ }\href {http://dx.doi.org/10.1038/nphys1342} {\ \textbf
		{\bibinfo {volume} {5}},\ \bibinfo {pages} {633 EP } (\bibinfo {year}
		{2009})}\BibitemShut {NoStop}%
	\bibitem [{\citenamefont {Reiter}\ \emph {et~al.}(2013)\citenamefont {Reiter},
		\citenamefont {Tornberg}, \citenamefont {Johansson},\ and\ \citenamefont
		{S\o{}rensen}}]{reiter_2013}%
	\BibitemOpen
	\bibfield  {author} {\bibinfo {author} {\bibfnamefont {F.}~\bibnamefont
			{Reiter}}, \bibinfo {author} {\bibfnamefont {L.}~\bibnamefont {Tornberg}},
		\bibinfo {author} {\bibfnamefont {G.}~\bibnamefont {Johansson}}, \ and\
		\bibinfo {author} {\bibfnamefont {A.~S.}\ \bibnamefont {S\o{}rensen}},\
	}\href {\doibase 10.1103/PhysRevA.88.032317} {\bibfield  {journal} {\bibinfo
		{journal} {Phys. Rev. A}\ }\textbf {\bibinfo {volume} {88}},\ \bibinfo
	{pages} {032317} (\bibinfo {year} {2013})}\BibitemShut {NoStop}%
\bibitem [{\citenamefont {Barreiro}\ \emph {et~al.}(2011)\citenamefont
	{Barreiro}, \citenamefont {M{\"u}ller}, \citenamefont {Schindler},
	\citenamefont {Nigg}, \citenamefont {Monz}, \citenamefont {Chwalla},
	\citenamefont {Hennrich}, \citenamefont {Roos}, \citenamefont {Zoller},\ and\
	\citenamefont {Blatt}}]{barreiro_2011}%
\BibitemOpen
\bibfield  {author} {\bibinfo {author} {\bibfnamefont {J.~T.}\ \bibnamefont
		{Barreiro}}, \bibinfo {author} {\bibfnamefont {M.}~\bibnamefont
		{M{\"u}ller}}, \bibinfo {author} {\bibfnamefont {P.}~\bibnamefont
		{Schindler}}, \bibinfo {author} {\bibfnamefont {D.}~\bibnamefont {Nigg}},
	\bibinfo {author} {\bibfnamefont {T.}~\bibnamefont {Monz}}, \bibinfo {author}
	{\bibfnamefont {M.}~\bibnamefont {Chwalla}}, \bibinfo {author} {\bibfnamefont
		{M.}~\bibnamefont {Hennrich}}, \bibinfo {author} {\bibfnamefont {C.~F.}\
		\bibnamefont {Roos}}, \bibinfo {author} {\bibfnamefont {P.}~\bibnamefont
		{Zoller}}, \ and\ \bibinfo {author} {\bibfnamefont {R.}~\bibnamefont
		{Blatt}},\ }\href {http://dx.doi.org/10.1038/nature09801} {\bibfield
	{journal} {\bibinfo  {journal} {Nature}\ }\textbf {\bibinfo {volume} {470}},\
	\bibinfo {pages} {486 EP } (\bibinfo {year} {2011})}\BibitemShut {NoStop}%
\bibitem [{\citenamefont {Lin}\ \emph {et~al.}(2013)\citenamefont {Lin},
	\citenamefont {Gaebler}, \citenamefont {Reiter}, \citenamefont {Tan},
	\citenamefont {Bowler}, \citenamefont {S{\o}rensen}, \citenamefont
	{Leibfried},\ and\ \citenamefont {Wineland}}]{lin_2013}%
\BibitemOpen
\bibfield  {author} {\bibinfo {author} {\bibfnamefont {Y.}~\bibnamefont
		{Lin}}, \bibinfo {author} {\bibfnamefont {J.~P.}\ \bibnamefont {Gaebler}},
	\bibinfo {author} {\bibfnamefont {F.}~\bibnamefont {Reiter}}, \bibinfo
	{author} {\bibfnamefont {T.~R.}\ \bibnamefont {Tan}}, \bibinfo {author}
	{\bibfnamefont {R.}~\bibnamefont {Bowler}}, \bibinfo {author} {\bibfnamefont
		{A.~S.}\ \bibnamefont {S{\o}rensen}}, \bibinfo {author} {\bibfnamefont
		{D.}~\bibnamefont {Leibfried}}, \ and\ \bibinfo {author} {\bibfnamefont
		{D.~J.}\ \bibnamefont {Wineland}},\ }\href
{http://dx.doi.org/10.1038/nature12801} {\ \textbf {\bibinfo {volume}
		{504}},\ \bibinfo {pages} {415 EP } (\bibinfo {year} {2013})}\BibitemShut
{NoStop}%
\bibitem [{\citenamefont {Kienzler}\ \emph {et~al.}(2015)\citenamefont
	{Kienzler}, \citenamefont {Lo}, \citenamefont {Keitch}, \citenamefont
	{de~Clercq}, \citenamefont {Leupold}, \citenamefont {Lindenfelser},
	\citenamefont {Marinelli}, \citenamefont {Negnevitsky},\ and\ \citenamefont
	{Home}}]{kienzler_2015}%
\BibitemOpen
\bibfield  {author} {\bibinfo {author} {\bibfnamefont {D.}~\bibnamefont
		{Kienzler}}, \bibinfo {author} {\bibfnamefont {H.-Y.}\ \bibnamefont {Lo}},
	\bibinfo {author} {\bibfnamefont {B.}~\bibnamefont {Keitch}}, \bibinfo
	{author} {\bibfnamefont {L.}~\bibnamefont {de~Clercq}}, \bibinfo {author}
	{\bibfnamefont {F.}~\bibnamefont {Leupold}}, \bibinfo {author} {\bibfnamefont
		{F.}~\bibnamefont {Lindenfelser}}, \bibinfo {author} {\bibfnamefont
		{M.}~\bibnamefont {Marinelli}}, \bibinfo {author} {\bibfnamefont
		{V.}~\bibnamefont {Negnevitsky}}, \ and\ \bibinfo {author} {\bibfnamefont
		{J.~P.}\ \bibnamefont {Home}},\ }\href {\doibase 10.1126/science.1261033}
{\bibfield  {journal} {\bibinfo  {journal} {Science}\ }\textbf {\bibinfo
		{volume} {347}},\ \bibinfo {pages} {53} (\bibinfo {year} {2015})},\ \Eprint
{http://arxiv.org/abs/http://science.sciencemag.org/content/347/6217/53.full.pdf}
{http://science.sciencemag.org/content/347/6217/53.full.pdf} \BibitemShut
{NoStop}%
\bibitem [{\citenamefont {Krauter}\ \emph {et~al.}(2011)\citenamefont
	{Krauter}, \citenamefont {Muschik}, \citenamefont {Jensen}, \citenamefont
	{Wasilewski}, \citenamefont {Petersen}, \citenamefont {Cirac},\ and\
	\citenamefont {Polzik}}]{krauter_2011}%
\BibitemOpen
\bibfield  {author} {\bibinfo {author} {\bibfnamefont {H.}~\bibnamefont
		{Krauter}}, \bibinfo {author} {\bibfnamefont {C.~A.}\ \bibnamefont
		{Muschik}}, \bibinfo {author} {\bibfnamefont {K.}~\bibnamefont {Jensen}},
	\bibinfo {author} {\bibfnamefont {W.}~\bibnamefont {Wasilewski}}, \bibinfo
	{author} {\bibfnamefont {J.~M.}\ \bibnamefont {Petersen}}, \bibinfo {author}
	{\bibfnamefont {J.~I.}\ \bibnamefont {Cirac}}, \ and\ \bibinfo {author}
	{\bibfnamefont {E.~S.}\ \bibnamefont {Polzik}},\ }\href {\doibase
	10.1103/PhysRevLett.107.080503} {\bibfield  {journal} {\bibinfo  {journal}
		{Phys. Rev. Lett.}\ }\textbf {\bibinfo {volume} {107}},\ \bibinfo {pages}
	{080503} (\bibinfo {year} {2011})}\BibitemShut {NoStop}%
\bibitem [{\citenamefont {Shankar}\ \emph {et~al.}(2013)\citenamefont
	{Shankar}, \citenamefont {Hatridge}, \citenamefont {Leghtas}, \citenamefont
	{Sliwa}, \citenamefont {Narla}, \citenamefont {Vool}, \citenamefont {Girvin},
	\citenamefont {Frunzio}, \citenamefont {Mirrahimi},\ and\ \citenamefont
	{Devoret}}]{shankar_2013}%
\BibitemOpen
\bibfield  {author} {\bibinfo {author} {\bibfnamefont {S.}~\bibnamefont
		{Shankar}}, \bibinfo {author} {\bibfnamefont {M.}~\bibnamefont {Hatridge}},
	\bibinfo {author} {\bibfnamefont {Z.}~\bibnamefont {Leghtas}}, \bibinfo
	{author} {\bibfnamefont {K.~M.}\ \bibnamefont {Sliwa}}, \bibinfo {author}
	{\bibfnamefont {A.}~\bibnamefont {Narla}}, \bibinfo {author} {\bibfnamefont
		{U.}~\bibnamefont {Vool}}, \bibinfo {author} {\bibfnamefont {S.~M.}\
		\bibnamefont {Girvin}}, \bibinfo {author} {\bibfnamefont {L.}~\bibnamefont
		{Frunzio}}, \bibinfo {author} {\bibfnamefont {M.}~\bibnamefont {Mirrahimi}},
	\ and\ \bibinfo {author} {\bibfnamefont {M.~H.}\ \bibnamefont {Devoret}},\
}\href {http://dx.doi.org/10.1038/nature12802} {\ \textbf {\bibinfo {volume}
	{504}},\ \bibinfo {pages} {419 EP } (\bibinfo {year} {2013})}\BibitemShut
{NoStop}%
\bibitem [{\citenamefont {Leghtas}\ \emph {et~al.}(2013)\citenamefont
	{Leghtas}, \citenamefont {Vool}, \citenamefont {Shankar}, \citenamefont
	{Hatridge}, \citenamefont {Girvin}, \citenamefont {Devoret},\ and\
	\citenamefont {Mirrahimi}}]{leghtas_2013}%
\BibitemOpen
\bibfield  {author} {\bibinfo {author} {\bibfnamefont {Z.}~\bibnamefont
		{Leghtas}}, \bibinfo {author} {\bibfnamefont {U.}~\bibnamefont {Vool}},
	\bibinfo {author} {\bibfnamefont {S.}~\bibnamefont {Shankar}}, \bibinfo
	{author} {\bibfnamefont {M.}~\bibnamefont {Hatridge}}, \bibinfo {author}
	{\bibfnamefont {S.~M.}\ \bibnamefont {Girvin}}, \bibinfo {author}
	{\bibfnamefont {M.~H.}\ \bibnamefont {Devoret}}, \ and\ \bibinfo {author}
	{\bibfnamefont {M.}~\bibnamefont {Mirrahimi}},\ }\href {\doibase
	10.1103/PhysRevA.88.023849} {\bibfield  {journal} {\bibinfo  {journal} {Phys.
			Rev. A}\ }\textbf {\bibinfo {volume} {88}},\ \bibinfo {pages} {023849}
	(\bibinfo {year} {2013})}\BibitemShut {NoStop}%
\bibitem [{\citenamefont {Kimchi-Schwartz}\ \emph {et~al.}(2016)\citenamefont
	{Kimchi-Schwartz}, \citenamefont {Martin}, \citenamefont {Flurin},
	\citenamefont {Aron}, \citenamefont {Kulkarni}, \citenamefont {Tureci},\ and\
	\citenamefont {Siddiqi}}]{kimchi_2016}%
\BibitemOpen
\bibfield  {author} {\bibinfo {author} {\bibfnamefont {M.~E.}\ \bibnamefont
		{Kimchi-Schwartz}}, \bibinfo {author} {\bibfnamefont {L.}~\bibnamefont
		{Martin}}, \bibinfo {author} {\bibfnamefont {E.}~\bibnamefont {Flurin}},
	\bibinfo {author} {\bibfnamefont {C.}~\bibnamefont {Aron}}, \bibinfo {author}
	{\bibfnamefont {M.}~\bibnamefont {Kulkarni}}, \bibinfo {author}
	{\bibfnamefont {H.~E.}\ \bibnamefont {Tureci}}, \ and\ \bibinfo {author}
	{\bibfnamefont {I.}~\bibnamefont {Siddiqi}},\ }\href {\doibase
	10.1103/PhysRevLett.116.240503} {\bibfield  {journal} {\bibinfo  {journal}
		{Phys. Rev. Lett.}\ }\textbf {\bibinfo {volume} {116}},\ \bibinfo {pages}
	{240503} (\bibinfo {year} {2016})}\BibitemShut {NoStop}%
\bibitem [{\citenamefont {DiCarlo}\ \emph {et~al.}(2009)\citenamefont
	{DiCarlo}, \citenamefont {Chow}, \citenamefont {Gambetta}, \citenamefont
	{Bishop}, \citenamefont {Johnson}, \citenamefont {Schuster}, \citenamefont
	{Majer}, \citenamefont {Blais}, \citenamefont {Frunzio}, \citenamefont
	{Girvin},\ and\ \citenamefont {Schoelkopf}}]{dicarlo_2009}%
\BibitemOpen
\bibfield  {author} {\bibinfo {author} {\bibfnamefont {L.}~\bibnamefont
		{DiCarlo}}, \bibinfo {author} {\bibfnamefont {J.~M.}\ \bibnamefont {Chow}},
	\bibinfo {author} {\bibfnamefont {J.~M.}\ \bibnamefont {Gambetta}}, \bibinfo
	{author} {\bibfnamefont {L.~S.}\ \bibnamefont {Bishop}}, \bibinfo {author}
	{\bibfnamefont {B.~R.}\ \bibnamefont {Johnson}}, \bibinfo {author}
	{\bibfnamefont {D.~I.}\ \bibnamefont {Schuster}}, \bibinfo {author}
	{\bibfnamefont {J.}~\bibnamefont {Majer}}, \bibinfo {author} {\bibfnamefont
		{A.}~\bibnamefont {Blais}}, \bibinfo {author} {\bibfnamefont
		{L.}~\bibnamefont {Frunzio}}, \bibinfo {author} {\bibfnamefont {S.~M.}\
		\bibnamefont {Girvin}}, \ and\ \bibinfo {author} {\bibfnamefont {R.~J.}\
		\bibnamefont {Schoelkopf}},\ }\href {http://dx.doi.org/10.1038/nature08121}
{\bibfield  {journal} {\bibinfo  {journal} {Nature}\ }\textbf {\bibinfo
		{volume} {460}},\ \bibinfo {pages} {240 EP } (\bibinfo {year}
	{2009})}\BibitemShut {NoStop}%
\bibitem [{\citenamefont {Rist{\`e}}\ \emph {et~al.}(2013)\citenamefont
	{Rist{\`e}}, \citenamefont {Dukalski}, \citenamefont {Watson}, \citenamefont
	{de~Lange}, \citenamefont {Tiggelman}, \citenamefont {Blanter}, \citenamefont
	{Lehnert}, \citenamefont {Schouten},\ and\ \citenamefont
	{DiCarlo}}]{riste_2013}%
\BibitemOpen
\bibfield  {author} {\bibinfo {author} {\bibfnamefont {D.}~\bibnamefont
		{Rist{\`e}}}, \bibinfo {author} {\bibfnamefont {M.}~\bibnamefont {Dukalski}},
	\bibinfo {author} {\bibfnamefont {C.~A.}\ \bibnamefont {Watson}}, \bibinfo
	{author} {\bibfnamefont {G.}~\bibnamefont {de~Lange}}, \bibinfo {author}
	{\bibfnamefont {M.~J.}\ \bibnamefont {Tiggelman}}, \bibinfo {author}
	{\bibfnamefont {Y.~M.}\ \bibnamefont {Blanter}}, \bibinfo {author}
	{\bibfnamefont {K.~W.}\ \bibnamefont {Lehnert}}, \bibinfo {author}
	{\bibfnamefont {R.~N.}\ \bibnamefont {Schouten}}, \ and\ \bibinfo {author}
	{\bibfnamefont {L.}~\bibnamefont {DiCarlo}},\ }\href
{http://dx.doi.org/10.1038/nature12513} {\bibfield  {journal} {\bibinfo
		{journal} {Nature}\ }\textbf {\bibinfo {volume} {502}},\ \bibinfo {pages}
	{350 EP } (\bibinfo {year} {2013})}\BibitemShut {NoStop}%
\bibitem [{\citenamefont {Roch}\ \emph {et~al.}(2014)\citenamefont {Roch},
	\citenamefont {Schwartz}, \citenamefont {Motzoi}, \citenamefont {Macklin},
	\citenamefont {Vijay}, \citenamefont {Eddins}, \citenamefont {Korotkov},
	\citenamefont {Whaley}, \citenamefont {Sarovar},\ and\ \citenamefont
	{Siddiqi}}]{roch_2014}%
\BibitemOpen
\bibfield  {author} {\bibinfo {author} {\bibfnamefont {N.}~\bibnamefont
		{Roch}}, \bibinfo {author} {\bibfnamefont {M.~E.}\ \bibnamefont {Schwartz}},
	\bibinfo {author} {\bibfnamefont {F.}~\bibnamefont {Motzoi}}, \bibinfo
	{author} {\bibfnamefont {C.}~\bibnamefont {Macklin}}, \bibinfo {author}
	{\bibfnamefont {R.}~\bibnamefont {Vijay}}, \bibinfo {author} {\bibfnamefont
		{A.~W.}\ \bibnamefont {Eddins}}, \bibinfo {author} {\bibfnamefont {A.~N.}\
		\bibnamefont {Korotkov}}, \bibinfo {author} {\bibfnamefont {K.~B.}\
		\bibnamefont {Whaley}}, \bibinfo {author} {\bibfnamefont {M.}~\bibnamefont
		{Sarovar}}, \ and\ \bibinfo {author} {\bibfnamefont {I.}~\bibnamefont
		{Siddiqi}},\ }\href {\doibase 10.1103/PhysRevLett.112.170501} {\bibfield
	{journal} {\bibinfo  {journal} {Phys. Rev. Lett.}\ }\textbf {\bibinfo
		{volume} {112}},\ \bibinfo {pages} {170501} (\bibinfo {year}
	{2014})}\BibitemShut {NoStop}%
\bibitem [{\citenamefont {Chantasri}\ \emph {et~al.}(2016)\citenamefont
	{Chantasri}, \citenamefont {E.~Kimchi-Schwartz}, \citenamefont {Roch},
	\citenamefont {Siddiqi},\ and\ \citenamefont {Jordan}}]{chantasri_2016}%
\BibitemOpen
\bibfield  {author} {\bibinfo {author} {\bibfnamefont {A.}~\bibnamefont
		{Chantasri}}, \bibinfo {author} {\bibfnamefont {M.}~\bibnamefont
		{E.~Kimchi-Schwartz}}, \bibinfo {author} {\bibfnamefont {N.}~\bibnamefont
		{Roch}}, \bibinfo {author} {\bibfnamefont {I.}~\bibnamefont {Siddiqi}}, \
	and\ \bibinfo {author} {\bibfnamefont {A.}~\bibnamefont {Jordan}},\
}\bibfield  {booktitle} {\emph {\bibinfo {booktitle} {Physical Review X}},\
}\href@noop {} {\ \textbf {\bibinfo {volume} {6}} (\bibinfo {year}
{2016})}\BibitemShut {NoStop}%
\bibitem [{\citenamefont {Liu}\ \emph {et~al.}(2016)\citenamefont {Liu},
	\citenamefont {Shankar}, \citenamefont {Ofek}, \citenamefont {Hatridge},
	\citenamefont {Narla}, \citenamefont {Sliwa}, \citenamefont {Frunzio},
	\citenamefont {Schoelkopf},\ and\ \citenamefont {Devoret}}]{liu_2016}%
\BibitemOpen
\bibfield  {author} {\bibinfo {author} {\bibfnamefont {Y.}~\bibnamefont
		{Liu}}, \bibinfo {author} {\bibfnamefont {S.}~\bibnamefont {Shankar}},
	\bibinfo {author} {\bibfnamefont {N.}~\bibnamefont {Ofek}}, \bibinfo {author}
	{\bibfnamefont {M.}~\bibnamefont {Hatridge}}, \bibinfo {author}
	{\bibfnamefont {A.}~\bibnamefont {Narla}}, \bibinfo {author} {\bibfnamefont
		{K.~M.}\ \bibnamefont {Sliwa}}, \bibinfo {author} {\bibfnamefont
		{L.}~\bibnamefont {Frunzio}}, \bibinfo {author} {\bibfnamefont {R.~J.}\
		\bibnamefont {Schoelkopf}}, \ and\ \bibinfo {author} {\bibfnamefont {M.~H.}\
		\bibnamefont {Devoret}},\ }\href {\doibase 10.1103/PhysRevX.6.011022}
{\bibfield  {journal} {\bibinfo  {journal} {Phys. Rev. X}\ }\textbf {\bibinfo
		{volume} {6}},\ \bibinfo {pages} {011022} (\bibinfo {year}
	{2016})}\BibitemShut {NoStop}%
\bibitem [{\citenamefont {Grossmann}\ \emph {et~al.}(1991)\citenamefont
	{Grossmann}, \citenamefont {Dittrich}, \citenamefont {Jung},\ and\
	\citenamefont {H\"anggi}}]{grossmann_1991}%
\BibitemOpen
\bibfield  {author} {\bibinfo {author} {\bibfnamefont {F.}~\bibnamefont
		{Grossmann}}, \bibinfo {author} {\bibfnamefont {T.}~\bibnamefont {Dittrich}},
	\bibinfo {author} {\bibfnamefont {P.}~\bibnamefont {Jung}}, \ and\ \bibinfo
	{author} {\bibfnamefont {P.}~\bibnamefont {H\"anggi}},\ }\href {\doibase
	10.1103/PhysRevLett.67.516} {\bibfield  {journal} {\bibinfo  {journal} {Phys.
			Rev. Lett.}\ }\textbf {\bibinfo {volume} {67}},\ \bibinfo {pages} {516}
	(\bibinfo {year} {1991})}\BibitemShut {NoStop}%
\bibitem [{\citenamefont {Bloch~Immanuel}(2012)}]{bloch_2012}%
\BibitemOpen
\bibfield  {author} {\bibinfo {author} {\bibfnamefont {N.~S.}\ \bibnamefont
		{Bloch~Immanuel}, \bibfnamefont {Dalibard~Jean}},\ }\href {\doibase
	doi:10.1038/nphys2259} {\bibfield  {journal} {\bibinfo  {journal} {Nature
			Physics}\ }\textbf {\bibinfo {volume} {267}},\ \bibinfo {pages} {2259}
	(\bibinfo {year} {2012})}\BibitemShut {NoStop}%
\bibitem [{\citenamefont {Gagnon}\ \emph {et~al.}(2017)\citenamefont {Gagnon},
	\citenamefont {Fillion-Gourdeau}, \citenamefont {Dumont}, \citenamefont
	{Lefebvre},\ and\ \citenamefont {MacLean}}]{gagnon_2017}%
\BibitemOpen
\bibfield  {author} {\bibinfo {author} {\bibfnamefont {D.}~\bibnamefont
		{Gagnon}}, \bibinfo {author} {\bibfnamefont {F.}~\bibnamefont
		{Fillion-Gourdeau}}, \bibinfo {author} {\bibfnamefont {J.}~\bibnamefont
		{Dumont}}, \bibinfo {author} {\bibfnamefont {C.}~\bibnamefont {Lefebvre}}, \
	and\ \bibinfo {author} {\bibfnamefont {S.}~\bibnamefont {MacLean}},\ }\href
{\doibase 10.1103/PhysRevLett.119.053203} {\bibfield  {journal} {\bibinfo
		{journal} {Phys. Rev. Lett.}\ }\textbf {\bibinfo {volume} {119}},\ \bibinfo
	{pages} {053203} (\bibinfo {year} {2017})}\BibitemShut {NoStop}%
\bibitem [{\citenamefont {Oliver}\ \emph {et~al.}(2005)\citenamefont {Oliver},
	\citenamefont {Yu}, \citenamefont {Lee}, \citenamefont {Berggren},
	\citenamefont {Levitov},\ and\ \citenamefont {Orlando}}]{oliver_2005}%
\BibitemOpen
\bibfield  {author} {\bibinfo {author} {\bibfnamefont {W.~D.}\ \bibnamefont
		{Oliver}}, \bibinfo {author} {\bibfnamefont {Y.}~\bibnamefont {Yu}}, \bibinfo
	{author} {\bibfnamefont {J.~C.}\ \bibnamefont {Lee}}, \bibinfo {author}
	{\bibfnamefont {K.~K.}\ \bibnamefont {Berggren}}, \bibinfo {author}
	{\bibfnamefont {L.~S.}\ \bibnamefont {Levitov}}, \ and\ \bibinfo {author}
	{\bibfnamefont {T.~P.}\ \bibnamefont {Orlando}},\ }\href {\doibase
	10.1126/science.1119678} {\bibfield  {journal} {\bibinfo  {journal}
		{Science}\ }\textbf {\bibinfo {volume} {310}},\ \bibinfo {pages} {1653}
	(\bibinfo {year} {2005})},\ \Eprint
{http://arxiv.org/abs/http://science.sciencemag.org/content/310/5754/\\
	1653.full.pdf} {http://science.sciencemag.org/content/310/5754/\\
	1653.full.pdf} \BibitemShut {NoStop}%
\bibitem [{\citenamefont {Sillanp\"a\"a}\ \emph {et~al.}(2006)\citenamefont
	{Sillanp\"a\"a}, \citenamefont {Lehtinen}, \citenamefont {Paila},
	\citenamefont {Makhlin},\ and\ \citenamefont {Hakonen}}]{sillanpaa_2006}%
\BibitemOpen
\bibfield  {author} {\bibinfo {author} {\bibfnamefont {M.}~\bibnamefont
		{Sillanp\"a\"a}}, \bibinfo {author} {\bibfnamefont {T.}~\bibnamefont
		{Lehtinen}}, \bibinfo {author} {\bibfnamefont {A.}~\bibnamefont {Paila}},
	\bibinfo {author} {\bibfnamefont {Y.}~\bibnamefont {Makhlin}}, \ and\
	\bibinfo {author} {\bibfnamefont {P.}~\bibnamefont {Hakonen}},\ }\href
{\doibase 10.1103/PhysRevLett.96.187002} {\bibfield  {journal} {\bibinfo
		{journal} {Phys. Rev. Lett.}\ }\textbf {\bibinfo {volume} {96}},\ \bibinfo
	{pages} {187002} (\bibinfo {year} {2006})}\BibitemShut {NoStop}%
\bibitem [{\citenamefont {Berns}\ \emph {et~al.}(2006)\citenamefont {Berns},
	\citenamefont {Oliver}, \citenamefont {Valenzuela}, \citenamefont {Shytov},
	\citenamefont {Berggren}, \citenamefont {Levitov},\ and\ \citenamefont
	{Orlando}}]{berns_2006}%
\BibitemOpen
\bibfield  {author} {\bibinfo {author} {\bibfnamefont {D.~M.}\ \bibnamefont
		{Berns}}, \bibinfo {author} {\bibfnamefont {W.~D.}\ \bibnamefont {Oliver}},
	\bibinfo {author} {\bibfnamefont {S.~O.}\ \bibnamefont {Valenzuela}},
	\bibinfo {author} {\bibfnamefont {A.~V.}\ \bibnamefont {Shytov}}, \bibinfo
	{author} {\bibfnamefont {K.~K.}\ \bibnamefont {Berggren}}, \bibinfo {author}
	{\bibfnamefont {L.~S.}\ \bibnamefont {Levitov}}, \ and\ \bibinfo {author}
	{\bibfnamefont {T.~P.}\ \bibnamefont {Orlando}},\ }\href {\doibase
	10.1103/PhysRevLett.97.150502} {\bibfield  {journal} {\bibinfo  {journal}
		{Phys. Rev. Lett.}\ }\textbf {\bibinfo {volume} {97}},\ \bibinfo {pages}
	{150502} (\bibinfo {year} {2006})}\BibitemShut {NoStop}%
\bibitem [{\citenamefont {Rudner}\ \emph {et~al.}(2008)\citenamefont {Rudner},
	\citenamefont {Shytov}, \citenamefont {Levitov}, \citenamefont {Berns},
	\citenamefont {Oliver}, \citenamefont {Valenzuela},\ and\ \citenamefont
	{Orlando}}]{rudner_2008}%
\BibitemOpen
\bibfield  {author} {\bibinfo {author} {\bibfnamefont {M.~S.}\ \bibnamefont
		{Rudner}}, \bibinfo {author} {\bibfnamefont {A.~V.}\ \bibnamefont {Shytov}},
	\bibinfo {author} {\bibfnamefont {L.~S.}\ \bibnamefont {Levitov}}, \bibinfo
	{author} {\bibfnamefont {D.~M.}\ \bibnamefont {Berns}}, \bibinfo {author}
	{\bibfnamefont {W.~D.}\ \bibnamefont {Oliver}}, \bibinfo {author}
	{\bibfnamefont {S.~O.}\ \bibnamefont {Valenzuela}}, \ and\ \bibinfo {author}
	{\bibfnamefont {T.~P.}\ \bibnamefont {Orlando}},\ }\href {\doibase
	10.1103/PhysRevLett.101.190502} {\bibfield  {journal} {\bibinfo  {journal}
		{Phys. Rev. Lett.}\ }\textbf {\bibinfo {volume} {101}},\ \bibinfo {pages}
	{190502} (\bibinfo {year} {2008})}\BibitemShut {NoStop}%
\bibitem [{\citenamefont {Izmalkov}\ \emph {et~al.}(2008)\citenamefont
	{Izmalkov}, \citenamefont {van~der Ploeg}, \citenamefont {Shevchenko},
	\citenamefont {Grajcar}, \citenamefont {Il'ichev}, \citenamefont {H\"ubner},
	\citenamefont {Omelyanchouk},\ and\ \citenamefont {Meyer}}]{izmalkov_2008}%
\BibitemOpen
\bibfield  {author} {\bibinfo {author} {\bibfnamefont {A.}~\bibnamefont
		{Izmalkov}}, \bibinfo {author} {\bibfnamefont {S.~H.~W.}\ \bibnamefont
		{van~der Ploeg}}, \bibinfo {author} {\bibfnamefont {S.~N.}\ \bibnamefont
		{Shevchenko}}, \bibinfo {author} {\bibfnamefont {M.}~\bibnamefont {Grajcar}},
	\bibinfo {author} {\bibfnamefont {E.}~\bibnamefont {Il'ichev}}, \bibinfo
	{author} {\bibfnamefont {U.}~\bibnamefont {H\"ubner}}, \bibinfo {author}
	{\bibfnamefont {A.~N.}\ \bibnamefont {Omelyanchouk}}, \ and\ \bibinfo
	{author} {\bibfnamefont {H.-G.}\ \bibnamefont {Meyer}},\ }\href {\doibase
	10.1103/PhysRevLett.101.017003} {\bibfield  {journal} {\bibinfo  {journal}
		{Phys. Rev. Lett.}\ }\textbf {\bibinfo {volume} {101}},\ \bibinfo {pages}
	{017003} (\bibinfo {year} {2008})}\BibitemShut {NoStop}%
\bibitem [{\citenamefont {Shevchenko}\ \emph {et~al.}(2010)\citenamefont
	{Shevchenko}, \citenamefont {Ashhab},\ and\ \citenamefont
	{Nori}}]{shevchenko_2010}%
\BibitemOpen
\bibfield  {author} {\bibinfo {author} {\bibfnamefont {S.}~\bibnamefont
		{Shevchenko}}, \bibinfo {author} {\bibfnamefont {S.}~\bibnamefont {Ashhab}},
	\ and\ \bibinfo {author} {\bibfnamefont {F.}~\bibnamefont {Nori}},\ }\href
{\doibase http://dx.doi.org/10.1016/j.physrep.2010.03.002} {\bibfield
	{journal} {\bibinfo  {journal} {Physics Reports}\ }\textbf {\bibinfo {volume}
		{492}},\ \bibinfo {pages} {1 } (\bibinfo {year} {2010})}\BibitemShut
{NoStop}%
\bibitem [{\citenamefont {Wilson}\ \emph {et~al.}(2010)\citenamefont {Wilson},
	\citenamefont {Johansson}, \citenamefont {Duty}, \citenamefont {Persson},
	\citenamefont {Sandberg},\ and\ \citenamefont {Delsing}}]{wilson_2010}%
\BibitemOpen
\bibfield  {author} {\bibinfo {author} {\bibfnamefont {C.~M.}\ \bibnamefont
		{Wilson}}, \bibinfo {author} {\bibfnamefont {G.}~\bibnamefont {Johansson}},
	\bibinfo {author} {\bibfnamefont {T.}~\bibnamefont {Duty}}, \bibinfo {author}
	{\bibfnamefont {F.}~\bibnamefont {Persson}}, \bibinfo {author} {\bibfnamefont
		{M.}~\bibnamefont {Sandberg}}, \ and\ \bibinfo {author} {\bibfnamefont
		{P.}~\bibnamefont {Delsing}},\ }\href {\doibase 10.1103/PhysRevB.81.024520}
{\bibfield  {journal} {\bibinfo  {journal} {Phys. Rev. B}\ }\textbf {\bibinfo
		{volume} {81}},\ \bibinfo {pages} {024520} (\bibinfo {year}
	{2010})}\BibitemShut {NoStop}%
\bibitem [{\citenamefont {Dupont-Ferrier}\ \emph {et~al.}(2013)\citenamefont
	{Dupont-Ferrier}, \citenamefont {Roche}, \citenamefont {Voisin},
	\citenamefont {Jehl}, \citenamefont {Wacquez}, \citenamefont {Vinet},
	\citenamefont {Sanquer},\ and\ \citenamefont {De~Franceschi}}]{dupont_2013}%
\BibitemOpen
\bibfield  {author} {\bibinfo {author} {\bibfnamefont {E.}~\bibnamefont
		{Dupont-Ferrier}}, \bibinfo {author} {\bibfnamefont {B.}~\bibnamefont
		{Roche}}, \bibinfo {author} {\bibfnamefont {B.}~\bibnamefont {Voisin}},
	\bibinfo {author} {\bibfnamefont {X.}~\bibnamefont {Jehl}}, \bibinfo {author}
	{\bibfnamefont {R.}~\bibnamefont {Wacquez}}, \bibinfo {author} {\bibfnamefont
		{M.}~\bibnamefont {Vinet}}, \bibinfo {author} {\bibfnamefont
		{M.}~\bibnamefont {Sanquer}}, \ and\ \bibinfo {author} {\bibfnamefont
		{S.}~\bibnamefont {De~Franceschi}},\ }\href {\doibase
	10.1103/PhysRevLett.110.136802} {\bibfield  {journal} {\bibinfo  {journal}
		{Phys. Rev. Lett.}\ }\textbf {\bibinfo {volume} {110}},\ \bibinfo {pages}
	{136802} (\bibinfo {year} {2013})}\BibitemShut {NoStop}%
\bibitem [{\citenamefont {Forster}\ \emph {et~al.}(2014)\citenamefont
	{Forster}, \citenamefont {Petersen}, \citenamefont {Manus}, \citenamefont
	{H\"anggi}, \citenamefont {Schuh}, \citenamefont {Wegscheider}, \citenamefont
	{Kohler},\ and\ \citenamefont {Ludwig}}]{forster_2014}%
\BibitemOpen
\bibfield  {author} {\bibinfo {author} {\bibfnamefont {F.}~\bibnamefont
		{Forster}}, \bibinfo {author} {\bibfnamefont {G.}~\bibnamefont {Petersen}},
	\bibinfo {author} {\bibfnamefont {S.}~\bibnamefont {Manus}}, \bibinfo
	{author} {\bibfnamefont {P.}~\bibnamefont {H\"anggi}}, \bibinfo {author}
	{\bibfnamefont {D.}~\bibnamefont {Schuh}}, \bibinfo {author} {\bibfnamefont
		{W.}~\bibnamefont {Wegscheider}}, \bibinfo {author} {\bibfnamefont
		{S.}~\bibnamefont {Kohler}}, \ and\ \bibinfo {author} {\bibfnamefont
		{S.}~\bibnamefont {Ludwig}},\ }\href {\doibase
	10.1103/PhysRevLett.112.116803} {\bibfield  {journal} {\bibinfo  {journal}
		{Phys. Rev. Lett.}\ }\textbf {\bibinfo {volume} {112}},\ \bibinfo {pages}
	{116803} (\bibinfo {year} {2014})}\BibitemShut {NoStop}%
\bibitem [{\citenamefont {Neilinger}\ \emph {et~al.}(2016)\citenamefont
	{Neilinger}, \citenamefont {Shevchenko}, \citenamefont {Bog\'ar},
	\citenamefont {Reh\'ak}, \citenamefont {Oelsner}, \citenamefont {Karpov},
	\citenamefont {H\"ubner}, \citenamefont {Astafiev}, \citenamefont {Grajcar},\
	and\ \citenamefont {Il'ichev}}]{neilinger_2016}%
\BibitemOpen
\bibfield  {author} {\bibinfo {author} {\bibfnamefont {P.}~\bibnamefont
		{Neilinger}}, \bibinfo {author} {\bibfnamefont {S.~N.}\ \bibnamefont
		{Shevchenko}}, \bibinfo {author} {\bibfnamefont {J.}~\bibnamefont {Bog\'ar}},
	\bibinfo {author} {\bibfnamefont {M.}~\bibnamefont {Reh\'ak}}, \bibinfo
	{author} {\bibfnamefont {G.}~\bibnamefont {Oelsner}}, \bibinfo {author}
	{\bibfnamefont {D.~S.}\ \bibnamefont {Karpov}}, \bibinfo {author}
	{\bibfnamefont {U.}~\bibnamefont {H\"ubner}}, \bibinfo {author}
	{\bibfnamefont {O.}~\bibnamefont {Astafiev}}, \bibinfo {author}
	{\bibfnamefont {M.}~\bibnamefont {Grajcar}}, \ and\ \bibinfo {author}
	{\bibfnamefont {E.}~\bibnamefont {Il'ichev}},\ }\href {\doibase
	10.1103/PhysRevB.94.094519} {\bibfield  {journal} {\bibinfo  {journal} {Phys.
			Rev. B}\ }\textbf {\bibinfo {volume} {94}},\ \bibinfo {pages} {094519}
	(\bibinfo {year} {2016})}\BibitemShut {NoStop}%
\bibitem [{\citenamefont {Stace}\ \emph {et~al.}(2005)\citenamefont {Stace},
	\citenamefont {Doherty},\ and\ \citenamefont {Barrett}}]{stace_2005}%
\BibitemOpen
\bibfield  {author} {\bibinfo {author} {\bibfnamefont {T.~M.}\ \bibnamefont
		{Stace}}, \bibinfo {author} {\bibfnamefont {A.~C.}\ \bibnamefont {Doherty}},
	\ and\ \bibinfo {author} {\bibfnamefont {S.~D.}\ \bibnamefont {Barrett}},\
}\href {\doibase 10.1103/PhysRevLett.95.106801} {\bibfield  {journal}
{\bibinfo  {journal} {Phys. Rev. Lett.}\ }\textbf {\bibinfo {volume} {95}},\
\bibinfo {pages} {106801} (\bibinfo {year} {2005})}\BibitemShut {NoStop}%
\bibitem [{\citenamefont {Stace}\ \emph {et~al.}(2013)\citenamefont {Stace},
	\citenamefont {Doherty},\ and\ \citenamefont {Reilly}}]{stace_2013}%
\BibitemOpen
\bibfield  {author} {\bibinfo {author} {\bibfnamefont {T.~M.}\ \bibnamefont
		{Stace}}, \bibinfo {author} {\bibfnamefont {A.~C.}\ \bibnamefont {Doherty}},
	\ and\ \bibinfo {author} {\bibfnamefont {D.~J.}\ \bibnamefont {Reilly}},\
}\href {\doibase 10.1103/PhysRevLett.111.180602} {\bibfield  {journal}
{\bibinfo  {journal} {Phys. Rev. Lett.}\ }\textbf {\bibinfo {volume} {111}},\
\bibinfo {pages} {180602} (\bibinfo {year} {2013})}\BibitemShut {NoStop}%
\bibitem [{\citenamefont {Ferr\'on}\ \emph {et~al.}(2012)\citenamefont
	{Ferr\'on}, \citenamefont {Dom\'{\i}nguez},\ and\ \citenamefont
	{S\'anchez}}]{ferron_2012}%
\BibitemOpen
\bibfield  {author} {\bibinfo {author} {\bibfnamefont {A.}~\bibnamefont
		{Ferr\'on}}, \bibinfo {author} {\bibfnamefont {D.}~\bibnamefont
		{Dom\'{\i}nguez}}, \ and\ \bibinfo {author} {\bibfnamefont {M.~J.}\
		\bibnamefont {S\'anchez}},\ }\href {\doibase 10.1103/PhysRevLett.109.237005}
{\bibfield  {journal} {\bibinfo  {journal} {Phys. Rev. Lett.}\ }\textbf
	{\bibinfo {volume} {109}},\ \bibinfo {pages} {237005} (\bibinfo {year}
	{2012})}\BibitemShut {NoStop}%
\bibitem [{\citenamefont {Ferr\'on}\ \emph {et~al.}(2016)\citenamefont
	{Ferr\'on}, \citenamefont {Dom\'{\i}nguez},\ and\ \citenamefont
	{S\'anchez}}]{ferron_2016}%
\BibitemOpen
\bibfield  {author} {\bibinfo {author} {\bibfnamefont {A.}~\bibnamefont
		{Ferr\'on}}, \bibinfo {author} {\bibfnamefont {D.}~\bibnamefont
		{Dom\'{\i}nguez}}, \ and\ \bibinfo {author} {\bibfnamefont {M.~J.}\
		\bibnamefont {S\'anchez}},\ }\href {\doibase 10.1103/PhysRevB.93.064521}
{\bibfield  {journal} {\bibinfo  {journal} {Phys. Rev. B}\ }\textbf {\bibinfo
		{volume} {93}},\ \bibinfo {pages} {064521} (\bibinfo {year}
	{2016})}\BibitemShut {NoStop}%
\bibitem [{not()}]{nota}%
\BibitemOpen
\href@noop {} {}\bibinfo {note} {We obtain similar results as reported here
	when an interaction term $\sigma_z^{(1)}\sigma_z^{(2)}$ is considered instead
	of $\sigma^{(1)}_{+}\sigma^{(2)}_{-} +
	\sigma^{(1)}_{-}\sigma^{(2)}_{+}$.}\BibitemShut {Stop}%
\bibitem [{\citenamefont {Berkley}\ \emph {et~al.}(2003)\citenamefont
	{Berkley}, \citenamefont {Xu}, \citenamefont {Ramos}, \citenamefont {Gubrud},
	\citenamefont {Strauch}, \citenamefont {Johnson}, \citenamefont {Anderson},
	\citenamefont {Dragt}, \citenamefont {Lobb},\ and\ \citenamefont
	{Wellstood}}]{berkley_2003}%
\BibitemOpen
\bibfield  {author} {\bibinfo {author} {\bibfnamefont {A.~J.}\ \bibnamefont
		{Berkley}}, \bibinfo {author} {\bibfnamefont {H.}~\bibnamefont {Xu}},
	\bibinfo {author} {\bibfnamefont {R.~C.}\ \bibnamefont {Ramos}}, \bibinfo
	{author} {\bibfnamefont {M.~A.}\ \bibnamefont {Gubrud}}, \bibinfo {author}
	{\bibfnamefont {F.~W.}\ \bibnamefont {Strauch}}, \bibinfo {author}
	{\bibfnamefont {P.~R.}\ \bibnamefont {Johnson}}, \bibinfo {author}
	{\bibfnamefont {J.~R.}\ \bibnamefont {Anderson}}, \bibinfo {author}
	{\bibfnamefont {A.~J.}\ \bibnamefont {Dragt}}, \bibinfo {author}
	{\bibfnamefont {C.~J.}\ \bibnamefont {Lobb}}, \ and\ \bibinfo {author}
	{\bibfnamefont {F.~C.}\ \bibnamefont {Wellstood}},\ }\href {\doibase
	10.1126/science.1084528} {\bibfield  {journal} {\bibinfo  {journal}
		{Science}\ }\textbf {\bibinfo {volume} {300}},\ \bibinfo {pages} {1548}
	(\bibinfo {year} {2003})}\BibitemShut {NoStop}%
\bibitem [{\citenamefont {Izmalkov}\ \emph {et~al.}(2004)\citenamefont
	{Izmalkov}, \citenamefont {Grajcar}, \citenamefont {Il'ichev}, \citenamefont
	{Wagner}, \citenamefont {Meyer}, \citenamefont {Smirnov}, \citenamefont
	{Amin}, \citenamefont {van~den Brink},\ and\ \citenamefont
	{Zagoskin}}]{izmalkov_2004}%
\BibitemOpen
\bibfield  {author} {\bibinfo {author} {\bibfnamefont {A.}~\bibnamefont
		{Izmalkov}}, \bibinfo {author} {\bibfnamefont {M.}~\bibnamefont {Grajcar}},
	\bibinfo {author} {\bibfnamefont {E.}~\bibnamefont {Il'ichev}}, \bibinfo
	{author} {\bibfnamefont {T.}~\bibnamefont {Wagner}}, \bibinfo {author}
	{\bibfnamefont {H.-G.}\ \bibnamefont {Meyer}}, \bibinfo {author}
	{\bibfnamefont {A.~Y.}\ \bibnamefont {Smirnov}}, \bibinfo {author}
	{\bibfnamefont {M.~H.~S.}\ \bibnamefont {Amin}}, \bibinfo {author}
	{\bibfnamefont {A.~M.}\ \bibnamefont {van~den Brink}}, \ and\ \bibinfo
	{author} {\bibfnamefont {A.~M.}\ \bibnamefont {Zagoskin}},\ }\href {\doibase
	10.1103/PhysRevLett.93.037003} {\bibfield  {journal} {\bibinfo  {journal}
		{Phys. Rev. Lett.}\ }\textbf {\bibinfo {volume} {93}},\ \bibinfo {pages}
	{037003} (\bibinfo {year} {2004})}\BibitemShut {NoStop}%
\bibitem [{\citenamefont {Majer}\ \emph {et~al.}(2005)\citenamefont {Majer},
	\citenamefont {Paauw}, \citenamefont {ter Haar}, \citenamefont {Harmans},\
	and\ \citenamefont {Mooij}}]{majer_2005}%
\BibitemOpen
\bibfield  {author} {\bibinfo {author} {\bibfnamefont {J.~B.}\ \bibnamefont
		{Majer}}, \bibinfo {author} {\bibfnamefont {F.~G.}\ \bibnamefont {Paauw}},
	\bibinfo {author} {\bibfnamefont {A.~C.~J.}\ \bibnamefont {ter Haar}},
	\bibinfo {author} {\bibfnamefont {C.~J. P.~M.}\ \bibnamefont {Harmans}}, \
	and\ \bibinfo {author} {\bibfnamefont {J.~E.}\ \bibnamefont {Mooij}},\ }\href
{\doibase 10.1103/PhysRevLett.94.090501} {\bibfield  {journal} {\bibinfo
		{journal} {Phys. Rev. Lett.}\ }\textbf {\bibinfo {volume} {94}},\ \bibinfo
	{pages} {090501} (\bibinfo {year} {2005})}\BibitemShut {NoStop}%
\bibitem [{\citenamefont {Liu}\ \emph {et~al.}(2006)\citenamefont {Liu},
	\citenamefont {Wei}, \citenamefont {Tsai},\ and\ \citenamefont
	{Nori}}]{liu_2006}%
\BibitemOpen
\bibfield  {author} {\bibinfo {author} {\bibfnamefont {Y.-x.}\ \bibnamefont
		{Liu}}, \bibinfo {author} {\bibfnamefont {L.~F.}\ \bibnamefont {Wei}},
	\bibinfo {author} {\bibfnamefont {J.~S.}\ \bibnamefont {Tsai}}, \ and\
	\bibinfo {author} {\bibfnamefont {F.}~\bibnamefont {Nori}},\ }\href {\doibase
	10.1103/PhysRevLett.96.067003} {\bibfield  {journal} {\bibinfo  {journal}
		{Phys. Rev. Lett.}\ }\textbf {\bibinfo {volume} {96}},\ \bibinfo {pages}
	{067003} (\bibinfo {year} {2006})}\BibitemShut {NoStop}%
\bibitem [{\citenamefont {Zhang}\ \emph {et~al.}(2009)\citenamefont {Zhang},
	\citenamefont {Liu}, \citenamefont {Li}, \citenamefont {Tarn},\ and\
	\citenamefont {Nori}}]{zhang_2009}%
\BibitemOpen
\bibfield  {author} {\bibinfo {author} {\bibfnamefont {J.}~\bibnamefont
		{Zhang}}, \bibinfo {author} {\bibfnamefont {Y.-x.}\ \bibnamefont {Liu}},
	\bibinfo {author} {\bibfnamefont {C.-W.}\ \bibnamefont {Li}}, \bibinfo
	{author} {\bibfnamefont {T.-J.}\ \bibnamefont {Tarn}}, \ and\ \bibinfo
	{author} {\bibfnamefont {F.}~\bibnamefont {Nori}},\ }\href {\doibase
	10.1103/PhysRevA.79.052308} {\bibfield  {journal} {\bibinfo  {journal} {Phys.
			Rev. A}\ }\textbf {\bibinfo {volume} {79}},\ \bibinfo {pages} {052308}
	(\bibinfo {year} {2009})}\BibitemShut {NoStop}%
\bibitem [{\citenamefont {Weber}\ \emph {et~al.}(2017)\citenamefont {Weber},
	\citenamefont {Samach}, \citenamefont {Hover}, \citenamefont {Gustavsson},
	\citenamefont {Kim}, \citenamefont {Melville}, \citenamefont {Rosenberg},
	\citenamefont {Sears}, \citenamefont {Yan}, \citenamefont {Yoder},
	\citenamefont {Oliver},\ and\ \citenamefont {Kerman}}]{weber_2017}%
\BibitemOpen
\bibfield  {author} {\bibinfo {author} {\bibfnamefont {S.~J.}\ \bibnamefont
		{Weber}}, \bibinfo {author} {\bibfnamefont {G.~O.}\ \bibnamefont {Samach}},
	\bibinfo {author} {\bibfnamefont {D.}~\bibnamefont {Hover}}, \bibinfo
	{author} {\bibfnamefont {S.}~\bibnamefont {Gustavsson}}, \bibinfo {author}
	{\bibfnamefont {D.~K.}\ \bibnamefont {Kim}}, \bibinfo {author} {\bibfnamefont
		{A.}~\bibnamefont {Melville}}, \bibinfo {author} {\bibfnamefont
		{D.}~\bibnamefont {Rosenberg}}, \bibinfo {author} {\bibfnamefont {A.~P.}\
		\bibnamefont {Sears}}, \bibinfo {author} {\bibfnamefont {F.}~\bibnamefont
		{Yan}}, \bibinfo {author} {\bibfnamefont {J.~L.}\ \bibnamefont {Yoder}},
	\bibinfo {author} {\bibfnamefont {W.~D.}\ \bibnamefont {Oliver}}, \ and\
	\bibinfo {author} {\bibfnamefont {A.~J.}\ \bibnamefont {Kerman}},\ }\href
{\doibase 10.1103/PhysRevApplied.8.014004} {\bibfield  {journal} {\bibinfo
		{journal} {Phys. Rev. Applied}\ }\textbf {\bibinfo {volume} {8}},\ \bibinfo
	{pages} {014004} (\bibinfo {year} {2017})}\BibitemShut {NoStop}%
\bibitem [{\citenamefont {Shevchenko}\ \emph {et~al.}(2008)\citenamefont
	{Shevchenko}, \citenamefont {van~der Ploeg}, \citenamefont {Grajcar},
	\citenamefont {Il'ichev}, \citenamefont {Omelyanchouk},\ and\ \citenamefont
	{Meyer}}]{shevchenko_2008}%
\BibitemOpen
\bibfield  {author} {\bibinfo {author} {\bibfnamefont {S.~N.}\ \bibnamefont
		{Shevchenko}}, \bibinfo {author} {\bibfnamefont {S.~H.~W.}\ \bibnamefont
		{van~der Ploeg}}, \bibinfo {author} {\bibfnamefont {M.}~\bibnamefont
		{Grajcar}}, \bibinfo {author} {\bibfnamefont {E.}~\bibnamefont {Il'ichev}},
	\bibinfo {author} {\bibfnamefont {A.~N.}\ \bibnamefont {Omelyanchouk}}, \
	and\ \bibinfo {author} {\bibfnamefont {H.-G.}\ \bibnamefont {Meyer}},\ }\href
{\doibase 10.1103/PhysRevB.78.174527} {\bibfield  {journal} {\bibinfo
		{journal} {Phys. Rev. B}\ }\textbf {\bibinfo {volume} {78}},\ \bibinfo
	{pages} {174527} (\bibinfo {year} {2008})}\BibitemShut {NoStop}%
\bibitem [{\citenamefont {Il'ichev}\ \emph {et~al.}(2010)\citenamefont
	{Il'ichev}, \citenamefont {Shevchenko}, \citenamefont {van~der Ploeg},
	\citenamefont {Grajcar}, \citenamefont {Temchenko}, \citenamefont
	{Omelyanchouk},\ and\ \citenamefont {Meyer}}]{ilichev_2010}%
\BibitemOpen
\bibfield  {author} {\bibinfo {author} {\bibfnamefont {E.}~\bibnamefont
		{Il'ichev}}, \bibinfo {author} {\bibfnamefont {S.~N.}\ \bibnamefont
		{Shevchenko}}, \bibinfo {author} {\bibfnamefont {S.~H.~W.}\ \bibnamefont
		{van~der Ploeg}}, \bibinfo {author} {\bibfnamefont {M.}~\bibnamefont
		{Grajcar}}, \bibinfo {author} {\bibfnamefont {E.~A.}\ \bibnamefont
		{Temchenko}}, \bibinfo {author} {\bibfnamefont {A.~N.}\ \bibnamefont
		{Omelyanchouk}}, \ and\ \bibinfo {author} {\bibfnamefont {H.-G.}\
		\bibnamefont {Meyer}},\ }\href {\doibase 10.1103/PhysRevB.81.012506}
{\bibfield  {journal} {\bibinfo  {journal} {Phys. Rev. B}\ }\textbf {\bibinfo
		{volume} {81}},\ \bibinfo {pages} {012506} (\bibinfo {year}
	{2010})}\BibitemShut {NoStop}%
\bibitem [{\citenamefont {Satanin}\ \emph {et~al.}(2012)\citenamefont
	{Satanin}, \citenamefont {Denisenko}, \citenamefont {Ashhab},\ and\
	\citenamefont {Nori}}]{satanin_2012}%
\BibitemOpen
\bibfield  {author} {\bibinfo {author} {\bibfnamefont {A.~M.}\ \bibnamefont
		{Satanin}}, \bibinfo {author} {\bibfnamefont {M.~V.}\ \bibnamefont
		{Denisenko}}, \bibinfo {author} {\bibfnamefont {S.}~\bibnamefont {Ashhab}}, \
	and\ \bibinfo {author} {\bibfnamefont {F.}~\bibnamefont {Nori}},\ }\href
{\doibase 10.1103/PhysRevB.85.184524} {\bibfield  {journal} {\bibinfo
		{journal} {Phys. Rev. B}\ }\textbf {\bibinfo {volume} {85}},\ \bibinfo
	{pages} {184524} (\bibinfo {year} {2012})}\BibitemShut {NoStop}%
\bibitem [{\citenamefont {Temchenko}\ \emph {et~al.}(2011)\citenamefont
	{Temchenko}, \citenamefont {Shevchenko},\ and\ \citenamefont
	{Omelyanchouk}}]{temchenko_2011}%
\BibitemOpen
\bibfield  {author} {\bibinfo {author} {\bibfnamefont {E.}~\bibnamefont
		{Temchenko}}, \bibinfo {author} {\bibfnamefont {S.}~\bibnamefont
		{Shevchenko}}, \ and\ \bibinfo {author} {\bibfnamefont {A.}~\bibnamefont
		{Omelyanchouk}},\ }\href {\doibase 10.1103/PhysRevB.83.144507} {\bibfield
	{journal} {\bibinfo  {journal} {Phys. Rev. B}\ }\textbf {\bibinfo {volume}
		{83}},\ \bibinfo {pages} {144507} (\bibinfo {year} {2011})}\BibitemShut
{NoStop}%
\bibitem [{\citenamefont {Sauer}\ \emph {et~al.}(2012)\citenamefont {Sauer},
	\citenamefont {Mintert}, \citenamefont {Gneiting},\ and\ \citenamefont
	{Buchleitner}}]{sauer_2012}%
\BibitemOpen
\bibfield  {author} {\bibinfo {author} {\bibfnamefont {S.}~\bibnamefont
		{Sauer}}, \bibinfo {author} {\bibfnamefont {F.}~\bibnamefont {Mintert}},
	\bibinfo {author} {\bibfnamefont {C.}~\bibnamefont {Gneiting}}, \ and\
	\bibinfo {author} {\bibfnamefont {A.}~\bibnamefont {Buchleitner}},\
}\href@noop {} {\bibfield  {journal} {\bibinfo  {journal} {journal of Physics
		B: Atomic, Molecular and Optical Physics}\ }\textbf {\bibinfo {volume}
	{45}},\ \bibinfo {pages} {154011} (\bibinfo {year} {2012})}\BibitemShut
{NoStop}%
\bibitem [{\citenamefont {Gramajo}\ \emph {et~al.}(2017)\citenamefont
	{Gramajo}, \citenamefont {Dom{\'\i}nguez},\ and\ \citenamefont
	{S{\'a}nchez}}]{gramajo_2017}%
\BibitemOpen
\bibfield  {author} {\bibinfo {author} {\bibfnamefont {A.~L.}\ \bibnamefont
		{Gramajo}}, \bibinfo {author} {\bibfnamefont {D.}~\bibnamefont
		{Dom{\'\i}nguez}}, \ and\ \bibinfo {author} {\bibfnamefont {M.~J.}\
		\bibnamefont {S{\'a}nchez}},\ }\href {\doibase 10.1140/epjb/e2017-80563-y}
{\bibfield  {journal} {\bibinfo  {journal} {Eur. Phys. J. B}\ }\textbf
	{\bibinfo {volume} {90}},\ \bibinfo {pages} {255} (\bibinfo {year}
	{2017})}\BibitemShut {NoStop}%
\bibitem [{\citenamefont {Kohler}\ \emph {et~al.}(1997)\citenamefont {Kohler},
	\citenamefont {Dittrich},\ and\ \citenamefont {H\"anggi}}]{kohler_1997}%
\BibitemOpen
\bibfield  {author} {\bibinfo {author} {\bibfnamefont {S.}~\bibnamefont
		{Kohler}}, \bibinfo {author} {\bibfnamefont {T.}~\bibnamefont {Dittrich}}, \
	and\ \bibinfo {author} {\bibfnamefont {P.}~\bibnamefont {H\"anggi}},\ }\href
{\doibase 10.1103/PhysRevE.55.300} {\bibfield  {journal} {\bibinfo  {journal}
		{Phys. Rev. E}\ }\textbf {\bibinfo {volume} {55}},\ \bibinfo {pages} {300}
	(\bibinfo {year} {1997})}\BibitemShut {NoStop}%
\bibitem [{\citenamefont {Wootters}(1998)}]{wootters_1998}%
\BibitemOpen
\bibfield  {author} {\bibinfo {author} {\bibfnamefont {W.~K.}\ \bibnamefont
		{Wootters}},\ }\href {\doibase 10.1103/PhysRevLett.80.2245} {\bibfield
	{journal} {\bibinfo  {journal} {Phys. Rev. Lett.}\ }\textbf {\bibinfo
		{volume} {80}},\ \bibinfo {pages} {2245} (\bibinfo {year}
	{1998})}\BibitemShut {NoStop}%
\bibitem [{\citenamefont {Nakamura}\ \emph {et~al.}(2001)\citenamefont
	{Nakamura}, \citenamefont {Pashkin},\ and\ \citenamefont
	{Tsai}}]{nakamura_2001}%
\BibitemOpen
\bibfield  {author} {\bibinfo {author} {\bibfnamefont {Y.}~\bibnamefont
		{Nakamura}}, \bibinfo {author} {\bibfnamefont {Y.~A.}\ \bibnamefont
		{Pashkin}}, \ and\ \bibinfo {author} {\bibfnamefont {J.~S.}\ \bibnamefont
		{Tsai}},\ }\href {\doibase 10.1103/PhysRevLett.87.246601} {\bibfield
	{journal} {\bibinfo  {journal} {Phys. Rev. Lett.}\ }\textbf {\bibinfo
		{volume} {87}},\ \bibinfo {pages} {246601} (\bibinfo {year}
	{2001})}\BibitemShut {NoStop}%
\bibitem [{\citenamefont {Hausinger}\ and\ \citenamefont
	{Grifoni}(2010)}]{grifoni_2010}%
\BibitemOpen
\bibfield  {author} {\bibinfo {author} {\bibfnamefont {J.}~\bibnamefont
		{Hausinger}}\ and\ \bibinfo {author} {\bibfnamefont {M.}~\bibnamefont
		{Grifoni}},\ }\href {\doibase 10.1103/PhysRevA.81.022117} {\bibfield
	{journal} {\bibinfo  {journal} {Phys. Rev. A}\ }\textbf {\bibinfo {volume}
		{81}},\ \bibinfo {pages} {022117} (\bibinfo {year} {2010})}\BibitemShut
{NoStop}%
\end{thebibliography}

\begin{thebibliography}{1}
	\bibitem{shirley}J. H. Shirley, Phys. Rev.{\bf 138}, B979  (1965).
	\bibitem{grifoni-hanggi} M. Grifoni and P. H\"anggi, Phys. Rep. {\bf 304}, 229 (1998).
	\bibitem{grifonih} J. Hausinger and M. Grifoni, Phys. Rev. A 81, 022117 (2010).
	\bibitem{fds} A. Ferr\'on, D. Dom\'{\i}nguez and M. J. S\'anchez, Phys. Rev. B.
	{\bf 82},134522 (2010).
	\bibitem{hanggi}
	S. Kohler, T. Dittrich and P. H\"anggi, Phys. Rev. E. {\bf 55}, 300  (1997).
	S. Kohler, R. Utermann, P. H\"anggi, and T. Dittrich,
	Phys. Rev. E. {\bf 58}, 7219 (1998).
	\bibitem{breuer} H.-P. Breuer, W. Huber and F. Petruccione, Phys. Rev. E 61, 4883
	(2000).
	\bibitem{ketzmerich} D. W. Hone, R. Ketzmerick and W. Kohn,  Phys. Rev. E 79, 051129 (2009). 
	\bibitem{fds2} A. Ferr\'on, D. Dom\'{\i}nguez and M. J. S\'anchez, Phys. Rev. Lett.
	{\bf 109}, 237005 (2012).
	\bibitem{fazio1} S. Gasparinetti, P. Solinas, S. Pugnetti, R. Fazio, and J. Pekola
	Phys. Rev. Lett. 110 150403 (2013).
	\bibitem{solinas2}
	V. Gramich, S. Gasparinetti, P. Solinas, and J. Ankerhold
	Phys. Rev. Lett. 113, 027001 (2014).
	\bibitem{ferron_2016} A. Ferr\'on, D. Dom\'inguez, and M. J. S\'anchez, Phys. Rev. B 93, 064521 (2016).
	\bibitem{2} A. L. Gramajo, D. Dom\'inguez, and M. J. S\'anchez, this paper (2018).
	\bibitem{3} A. L. Gramajo, D. Dom\'inguez, and M. J. S\'anchez, Eur. Phys. J. B 90, 255 (2017).
	\bibitem{4} S.-K. Son, S. Han, and S.-I. Chu, Phys. Rev. A 79, 032301 (2009).
	
	
\end{thebibliography}
%

\widetext
\clearpage

\begin{center}
	\textbf{\Large Supplementary Information: \\ \medskip
		\Large Amplitude tuning of steady state entanglement in strongly driven coupled qubits }
\end{center}

\setcounter{equation}{0}
\setcounter{figure}{0}
\setcounter{table}{0}

\setcounter{page}{1}
\makeatletter
\renewcommand{\theequation}{S\arabic{equation}}
\renewcommand{\thefigure}{S\arabic{figure}}
\renewcommand{\bibnumfmt}[1]{[S#1]}
\renewcommand{\citenumfont}[1]{S#1}

\section{\label{sec:A} Floquet-Markov Master Equation}

\hspace*{0.6cm} The open system dynamics can be described by the global Hamiltonian 
\begin{equation}
\begin{aligned}
\hat{\cal H}(t)=\hat{H}_{s}(t) + \hat{H}_{b} + \hat{H}_{sb},
\label{eq:A1}
\end{aligned}
\end{equation} 
where $\hat{H}_{s}(t)=\hat{H}_{0} + \hat{V}(t)$ corresponds to  the Hamiltonian of two coupled qubits $H_0$ driving by periodic external fields $\hat{V}(t)$. Since 
$\hat{H}_{s}(t)=\hat{H}_{s}(t+ \tau )$ with  $\tau=2\pi/\omega_{0}$ the driving period, it is convenient to use the Floquet formalism,  that allows to treat  periodic forces
of arbitrary strength and frequency. \cite{shirley,grifoni-hanggi,grifonih,fds}
In  the Floquet formalism, 
the solutions of the   time dependent Schr\"odinger equation are of the
form  $|\Psi_\alpha(t)\rangle=e^{i\gamma_\alpha t/\hbar}|\alpha(t)\rangle$, where
the  Floquet states $|\alpha(t)\rangle$
satisfy $|\alpha(t)\rangle$=$|\alpha(t+ \tau)\rangle = 
\sum_K |u_\alpha(K) \rangle e^{-iK\omega t}$, and
are eigenstates of the equation
$[{\cal H} (t)- i \hbar \partial/\partial t ] |\alpha(t)\rangle= \gamma_\alpha |\alpha(t)\rangle$,
with $\gamma_\alpha$ the associated quasi-energy.

We consider a bosonic thermal bath at temperature $T_{b}$ described by the usual harmonic oscillators Hamiltonian $\hat{H}_{b}$, which is linearly coupled to  the two-qubits system in the form $\hat{H}_{sb}=g\mathcal{A}\otimes\mathcal{B}$, with $g$ the coupling strength, $\mathcal{B}$ an observable of the bath and $\mathcal{A}$ an observable of the system. The bath  degrees of freedom
are characterized by the  spectral density  $J(\Omega)=\gamma\Omega e^{-|\Omega|/\omega_{c}}$, with  $\omega_{c}$ the cutoff frequency. It is further assumed that at time $t=0$ the 
bath is in thermal equilibrium and uncorrelated with the system.

The global dynamics  obeys the Von-Neumann equation
\begin{equation}
\begin{aligned}
\dot{\rho}_{tot}(t)=-\frac{i}{\hbar} \left[\hat{\mathcal{H}}(t), \rho_{tot}(t)\right];
\label{eq:A2}
\end{aligned}
\end{equation} which after tracing over the degree of freedom of the bath becomes an equation for the  evolution of  the two-qubits  reduced density matrix $\rho(t)=Tr_{b}\left(  \rho_{tot} (t)\right)$,
\begin{equation}
\begin{aligned}
\dot{\rho}(t)=-\frac{i}{\hbar} Tr_{b} \left( \left[ \hat{\mathcal{H}}(t), \rho_{tot}(t)\right] \right). 
\label{eq:A3}
\end{aligned}
\end{equation}  
After expanding $\rho(t)$ in terms of  the time-periodic  Floquet basis $\{|{\alpha}(t)\rangle\}$, 
(${\alpha}, {\beta}=0,1,2,3$)
\begin{equation}
\rho_{\alpha\beta}(t)=\langle {\alpha}(t)|\rho(t)|{\beta}(t)\rangle\;,
\end{equation}
the Born (weak coupling) and Markov (fast relaxation) approximations for the time evolution  are performed.
In this way, the Floquet-Markov Master equation   \cite{grifoni-hanggi,grifonih,hanggi,breuer,ketzmerich,fds2,fazio1,solinas2} is 
obtained:
\begin{equation}
\begin{aligned}
\dot{\rho}_{\alpha\beta}(t)&=-i(\gamma_\alpha-\gamma_\beta)\rho_{\alpha\beta}-\sum_{\alpha'\beta'} \mathcal{L}_{\alpha\beta,\alpha'\beta'} (t)\rho(t)_{\alpha'\beta'},\\
\mathcal{L}_{\alpha\beta,\alpha'\beta'} (t)&=\sum_{Q}  \mathcal{L}^{Q}_{\alpha\beta,\alpha'\beta'} e^{-iQ\omega t},
\label{eq:A4}
\end{aligned}
\end{equation} with $ \mathcal{L}_{\alpha\beta,\alpha'\beta'} (t)$ the transition rates and $Q\in\mathbb{Z}$.  The Fourier coefficients are defined as
\begin{equation} 
\begin{aligned}
\mathcal{L}^{Q}_{\alpha\beta,\alpha'\beta'} &=  \sum_{K} \left( \delta_{\beta\beta'}\sum_{\eta} g^{K}_{\eta\alpha'} A^{K+Q}_{\alpha\eta}A^{K}_{\eta\alpha'}    +  \delta_{\alpha\alpha'}\sum_{\eta} g^{-K}_{\eta\beta'} A^{K+Q}_{\eta\beta}A^{K}_{\beta'\eta}    - \left(g^{K}_{\alpha\alpha'} + g^{-K-Q}_{\beta\beta'}\right) A^{K}_{\alpha\alpha'}A^{K+Q}_{\beta'\beta}\right),
\label{eq:A5}
\end{aligned}
\end{equation} with $g^{K}_{\alpha\beta} = J(\gamma_{\alpha\beta} + K\omega ) n_{th} (\gamma_{\alpha\beta} + K\omega)$, and $\gamma_{\alpha\beta} = \gamma_{\alpha} - \gamma_{\beta}$ ($\gamma_{\alpha}$ the quasienergy associated to the Floquet state $|u_{\alpha}(t)\rangle$) and $K\in\mathbb{Z}$.
The thermal occupation is given by the  Bose-Einstein function 
$n_{th}(x)=1/(e^{x/k_{B}T} - 1)$. 

Each $A^{K}_{\alpha\beta}$ is a transition matrix element in the Floquet basis, defined as $$A^{K}_{\alpha\beta}=\sum_{L} \langle u_{\alpha} (L)|\mathcal{A}|u_{\beta}(L+K)\rangle,$$ with $|u_{\alpha}(L)\rangle$ the  $L\in\mathbb{Z}$ Fourier component of the Floquet state.

Considering that the time scale $t_{r}$ for full relaxation satisfies $t_{r} \gg \tau$, the transition rates  $ \mathcal{L}_{\alpha\beta,\alpha'\beta'} (t)$  can be approximated by their average over one period $T$,  $ \mathcal{L}_{\alpha\beta,\alpha'\beta'} (t)\sim  {L}^{Q=0}_{\alpha\beta,\alpha'\beta'}$ \cite{kohler}, obtaining
\begin{equation}
\begin{aligned}
\mathcal{L}^{Q=0}_{\alpha\beta,\alpha'\beta'}= {L}_{\alpha\beta,\alpha'\beta'} =  \delta_{\beta\beta'} \sum_{\eta} {R}_{\eta\eta,\alpha'\alpha}    + \delta_{\alpha\alpha'}\sum_{\eta} ({R}_{\eta\eta,\beta'\beta})^{*}     - {R}_{\alpha\beta,\alpha'\beta'} - ({R}_{\beta\alpha,\beta'\alpha'})^{*},
\label{eq:A6}
\end{aligned}
\end{equation} where the rates
\begin{equation}
\begin{aligned}
{R}_{\alpha\beta,\alpha'\beta'} &= \sum_{Q} g_{\alpha\alpha'}^{Q}  A^{Q}_{\alpha\alpha'} \left( A^{Q}_{\beta\beta'}\right)^{*},
\label{eq:A7}
\end{aligned}
\end{equation}  can be interpreted as sums of Q-photon exchange terms.

We obtain numerically the Floquet components $|u_\alpha(K)\rangle$ and then we
calculate the rates  ${R}_{\alpha\beta\alpha'\beta'}$ and ${L}_{\alpha\beta\alpha'\beta'}$ . After obtaining the  ${L}_{\alpha\beta\alpha'\beta'}$ terms, the
time dependent solution of $\rho_{\alpha\beta}(t)$ and the steady state $\rho_{\alpha\beta}(t\rightarrow\infty)$   are computed as
described in Ref.\cite{ferron_2016}.

\section{\label{sec:B} Resonance conditions and quantum state tomography at resonances}
\hspace*{0.6cm} From the results presented along this  work \cite{2}, it follows that the relevant entanglement dynamics takes place near the resonance conditions. We classify the resonances according to the involved states: the SS-resonance (separable-separable states), SE-resonance (separable-entangled states and vice versa) and  EE-resonance (entangled-entangled states).

\begin{figure}[htb]
	\centering
	\includegraphics[scale = 0.38]{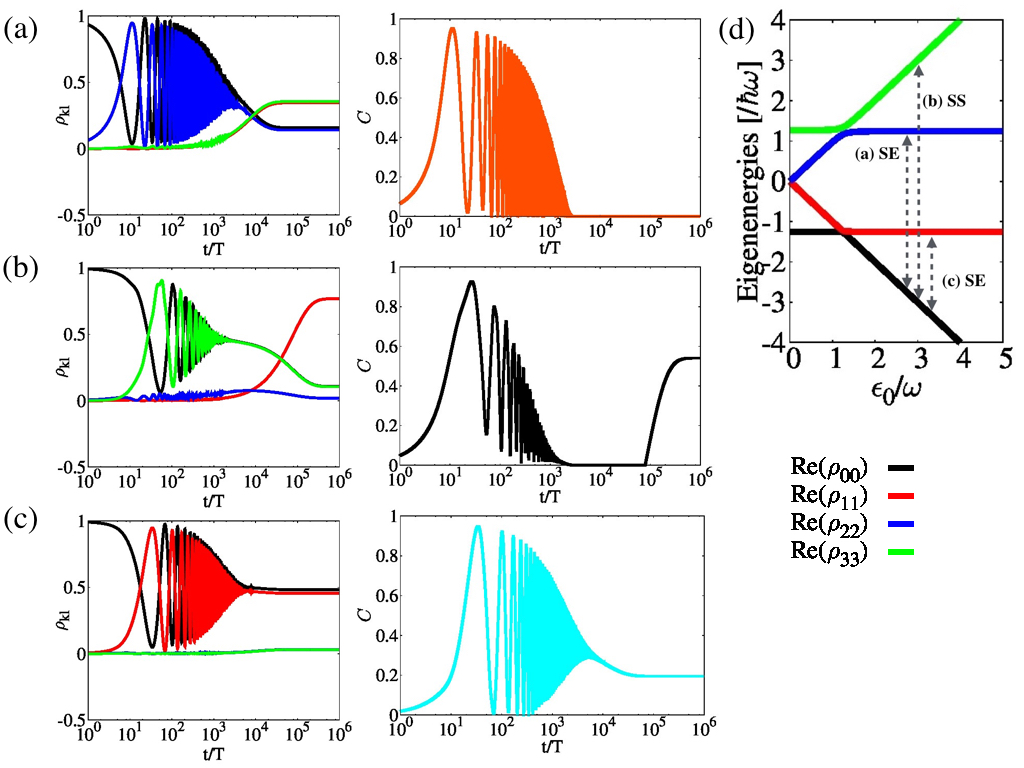}
	\caption{Left column displays the quantum tomography $\rho_{kl}$ as function of time $t/T$, in the eigenstates basis $\mathcal{H}=\{E_{k}\}$, $k=0,1,2,3$, for $\epsilon_{0}/\omega=2.75$(a), $\epsilon_{0}/\omega=3$(b) and $\epsilon_{0}/\omega =3.25$ (c). Central column presents the corresponding plots for the time evolution of concurrence $C$. The eigenenergies $E_{k}$  as function of $\epsilon_{0}/\omega$ are plotted in (d) together with a scheme of the different kind of resonances indicated by the dashed lines. All the cases correspond to $A/\omega=3.8$. The other parameters are:$\Delta_{2}/\Delta_1=1.5$ , $J/\Delta_1=-25$ and $\omega/\Delta_1=10$. The bath temperature is taken as $T_{b}/\Delta_1=0.0467$ and for the bath spectral density we consider $\gamma=0.001$ with a cutoff frequency $\omega_c/\Delta_1=333$. We choose as initial condition $|E_{0}\rangle$, corresponding to the ground state of the  undriven  Hamiltonian.} 
	\label{sfig:1}
\end{figure}

We calculate the resonance conditions considering the Hamiltonian  $H_s(t)=\hat{H}_{0} + \hat{V}(t)$, as it was already done in Ref.\cite{3}.

The  Hamiltonian is 
\begin{equation}
\hat{H}_{0}= -\sum_{i=1,2} \left( \epsilon_{0} \sigma_{z}^{(i)} + \Delta_{i}\sigma_{x}^{(i)} \right)/2 + \hat{H}_{12},
\end{equation}
with 
\begin{equation}
\hat{H}_{12}= -J/2\left(  \sigma_{+}^{(1)}\sigma_{-}^{(2)} + \sigma_{-}^{(1)}\sigma_{+}^{(2)} \right),
\end{equation}  
the interaction term and $J$ the coupling strength.  
The system Hamiltonian $H_0$, for $\Delta_i\ll\epsilon_0$, has two entangled eigenstates $|e_{\pm}\rangle\approx \frac{1}{\sqrt{2}}(|01\rangle \pm|10\rangle)$ (in the basis spanned by the eigenstates of  $\sigma^{(1)}_{z}\otimes\sigma^{(2)}_{z}$) with eigenenergies $E_{e\pm}\approx\mp J/2$, and two separable (disentangled) eigenstates $|s_0\rangle\approx|00\rangle$ and  $|s_1\rangle\approx|11\rangle$, with eigenenergies $E_{s0}\approx-\epsilon_0$ and $E_{s1}\approx\epsilon_0$, respectively.

The external ac field is 
$\hat{V}(t)=-A/2\cos(\omega t)\left(\sigma_{z}^{(1)} + \sigma_{z}^{(2)}\right)$, with $A$ the amplitude and $\omega$ the frequency of the driving. In the Floquet approach, the resonance conditions correspond to  $\gamma_{\alpha}-\gamma_{\beta}= n \omega$.
The quasienergies,  computed  for $\Delta_{i}/\omega\rightarrow 0$ using 
perturbation theory \cite{4} in the lowest order, are  $\gamma_{\alpha}\sim  \pm \epsilon_{0} + \tilde{m} \omega$ and $ \pm J/2 + \tilde{m} \omega$ with $\tilde{m}\in\mathbb{Z}$, which correspond to separable and entangled states, respectively. 
As, for $\Delta_{i}/\omega\rightarrow 0$,
the driving $\hat{V}(t)$ and the coupling Hamiltonian commute, the location of the avoided (quasi)
crossings in the spectrum of quasienergies are replicas (in $\pm n \omega$)   of the quasi crossings of  the  static spectrum. Therefore  the resonance conditions $\gamma_{\alpha}-\gamma_{\beta}= n \omega$, $n\in\mathbb{Z}$ \cite{4}  are thus satisfied respectively for $\epsilon_{0} \pm J/2 \sim m\omega $ (SE-resonances), $2\epsilon_{0}\sim m\omega $ (SS-resonances) and  $J\sim m\omega $  (EE-resonances) with $m\in\mathbb{Z}$. 
In Figure(\ref{sfig:1}) (d) we show the energy spectrum of $H_0$ as a function of $\epsilon_{0}$, and the location of some resonances are plotted with dashed black lines, allowing to identify each resonance condition.

The dynamics at each type of resonance can be visualized in terms of  quantum tomography, by evaluating the time evolution of individual components of the density matrix.
Figure(\ref{sfig:1}) shows the numerical results for the  time-evolution of the reduced density matrix elements 
${\rho_{kl}} (t)$ and concurrence $C$, with $k,l=0,...,3$, indexes corresponding to the eigenstates basis $\{E_{k}\}$ of  $\hat{H}_{0}$. Each plot corresponds to a different  resonance condition: $\epsilon_{0}/\omega=2.75$ (\ref{sfig:1}.a) (SE-resonance), $\epsilon_{0}/\omega=3$ (\ref{sfig:1}.b) (SS-resonance) and $\epsilon_{0}/\omega=3.25$ (\ref{sfig:1}.c) (SE-resonance).  In all the cases we see
that  there are Rabi-like oscillations of the populations of the two states involved in the resonances  inducing a dynamic generation of entanglement, for times $t\ll t_c$, with $t_c$ the decoherence time,.  Above $t_c$ the oscillations are damped, coherence is lost and there is no entanglement. Beyond the relaxation time $t_r$, a new regime sets in, as discussed in this work \cite{2}.

\section{\label{sec:C} Relaxation rates and population inversion processes}
\hspace*{0.6cm}As it was presented in \cite{2}, our system of work exhibits processes of population inversion in the stationary regime, due to the action of the driving and the system-bath interaction. In the case of two-level systems, these mechanisms can be cast in terms of virtual photon exchange processes with the bath, which contribute to the system relaxation rate   $\Gamma_{r} = \sum_{n}\Gamma_{r}^{(n)}$ \cite{ferron_2016}, $n\in{Z}$, where $\Gamma_{r}^{(0)}$ describes the conventional relaxation process (without exchange of virtual photon)  and $\Gamma_{r}^{(\pm n)}$ corresponds to the ac contribution due to the exchange of $n$ virtual photon with energy $\pm m \hbar \omega$. 
Thus, whenever there is population inversion, the dc relaxation terms vanish $\Gamma_{r}^{(0)}$ while
it is expected that one of the  $\Gamma_{r}^{(-m)}$ terms becomes the dominant one.


In this section, we  estimate the rates  $\Gamma_{r}^{(n)}$ near a multiphoton resonance after performing two approximations:

(i) As it was shown in the examples plotted in Figure(\ref{sfig:1}) , 
at and near a resonance the population is concentrated in  the two states intervening in the resonance. 
Therefore, we reduce the dynamics to the subspace of the two Floquet states $\{ |{a}(t)\rangle,|{b}(t)\rangle \}$
that satisfy the resonance condition $\gamma_{a}-\gamma_{b}\approx n \omega$.

(ii) In the secular approximation,  the density  matrix can be taken as approximately diagonal in the Floquet basis for times scales satisfying  $t\gg t_{c}$ (with $t_{c}$ the decoherence time) \cite{breuer,ketzmerich,fazio1}. 
This approximation fails exactly at the resonance \cite{ketzmerich,fazio1}, but works well off-resonance and even near a resonance for sufficiently small coupling to the environment \cite{fazio1,ferron_2016}.
We can thus approximate the matrix elements as $\rho_{\alpha\beta}\sim\rho_{\alpha\alpha}$ and $\rho_{\alpha\beta} \sim 0 $ for $\alpha\neq\beta$. Each $\rho_{\alpha\alpha}$ can be interpreted as $\rho_{\alpha\alpha}=P_{\alpha}$ with $P_{\alpha}$ the population  of the Floquet state $|{\alpha}(t)\rangle$.

After these two approximations, we can   write a Pauli  like equation for the populations near a multiphoton resonance:
\begin{eqnarray}
\frac{dP_\alpha}{dt}&=&\sum_{\beta} L_{\alpha\alpha\beta\beta} P_\beta\nonumber\\
&=& 2\sum_{\beta}R_{\alpha\alpha\beta\beta}P_\beta - R_{\beta\beta\alpha\alpha}P_\alpha
\end{eqnarray}
where $R_{\alpha\alpha\beta\beta}=\sum_n
g_{\alpha\beta}^n|A_{\alpha\beta}^n|^2$, from Eq.(\ref{eq:A7}).
For the two Floquet states $|a\rangle,|b\rangle$, the  relaxation  rate  that follows from the above equation is
$\Gamma_r = 2(R_{aabb}+R_{bbaa})$. 
Using Eq.(\ref{eq:A7}), we can
decompose the relaxation rate as a sum of  terms that
describe virtual  n-photon transitions:\cite{grifonih}
\begin{equation}
\Gamma_r = \Gamma_r^{(0)} +\sum_{n\not=0} \Gamma_r^{(n)}\;,
\end{equation}
with
\begin{equation}
\Gamma_r^{(n)}=2(g_{ab}^n|A_{ab}^n|^2+g_{ba}^n|A_{ba}^n|^2).
\label{eq:C2}
\end{equation}

\begin{figure}[!htb]
	\centering
	\includegraphics[scale = 0.5]{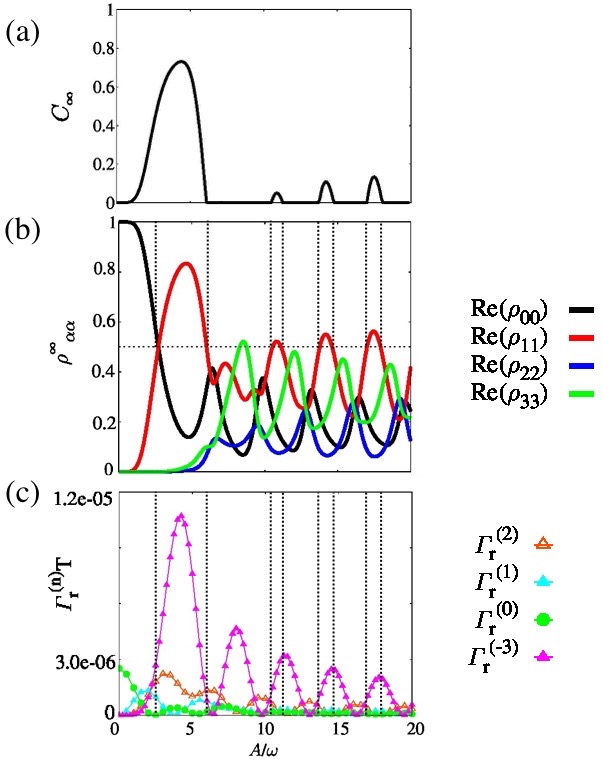}
	\caption{ Plot of  $C_{\infty}$ (a)  $\rho^{\infty}_{\alpha\alpha}$ (b) and  relevant $\Gamma_{r}^{(n)}$  terms that contribute to the relaxation rate (c)  as a function of $A/\omega$. All the  cases correspond to the fixed value $\epsilon_{0}/\omega=4.1$. The other parameters are the same presented in Fig.(\ref{sfig:1}).} 
	\label{sfig:2}
\end{figure}

As an example, we choose the off-resonant case $\epsilon_{0}/\omega=4.1$, that is close to the $m=3$-photon resonance at $\epsilon_{0}/\omega=4.25$.  We plot in  Fig.\ref{sfig:2} as a function of $A$ the concurrence, $C_\infty$, the population of the states in the steady state, $\rho^{\infty}_{\alpha\alpha}$, and the relaxation rates $\Gamma_r^{(n)} $. While the amplitude $A\lesssim|\epsilon_{0}|$ the population transfer only takes place between the states $|E_{0}\rangle$ (black line) and $|E_{1}\rangle$ (red line), see Fig.(\ref{sfig:1})(d). When the amplitude increases,  more avoided crossings are reached in the range $[\epsilon_0-A,\epsilon_0+A]$ spanned by the driving,   allowing  to  populate the states $|E_{2}\rangle$ and $|E_{3}\rangle$ .  In this later situation the two-level approximation losses validity for $A\gg |\epsilon_{0}|$.  For amplitudes $A$ within the two-level regime, we see that there is population inversion whenever the $\Gamma_{r}^{(-3)}$ term is the largest one and $C_{\infty}\not=0$. It is interesting to point out that  for large $A$, where the calculation of the plotted $\Gamma_{r}^{(n)}$ is no longer valid,  there is a finite value of concurrence whenever the population of the first excited state is the largest. Thus, even when the dynamics is more complex, involving the all the four states, the qualitative picture of entanglement generation due to population inversion is still correct.


\end{document}